\documentclass[pre,floatfix,twocolumn,showpacs]{revtex4}

\usepackage{amsmath}
\usepackage{amssymb}
\usepackage{graphicx}
\usepackage{psfrag}

\newcommand{\ie}{\emph{i.}$\,$\emph{e.}\ }
\newcommand{\eg}{\emph{e.}$\,$\emph{g.}\ }
\newcommand{\etal}{\emph{et}$\,$\emph{al.}}
\newcommand{\cf}{\emph{cf.}\ }

\newcommand{\romd}{{\text{d}}}

\newcommand{\romG}{{\text{G}}}

\newcommand{\VECalpha}{{\boldsymbol{\alpha}}}

\newcommand{\VECe}{{\boldsymbol{e}}}
\newcommand{\VECf}{{\boldsymbol{f}}}
\newcommand{\VECl}{{\boldsymbol{l}}}
\newcommand{\VECm}{{\boldsymbol{m}}}
\newcommand{\VECn}{{\boldsymbol{n}}}
\newcommand{\VECr}{{\boldsymbol{r}}}
\newcommand{\VECt}{{\boldsymbol{t}}}
\newcommand{\VECx}{{\boldsymbol{x}}}
\newcommand{\VECy}{{\boldsymbol{y}}}
\newcommand{\VECz}{{\boldsymbol{z}}}

\newcommand{\VECF}{{\boldsymbol{F}}}
\newcommand{\VECM}{{\boldsymbol{M}}}

\newcommand{\VECX}{{\boldsymbol{X}}}

\newcommand{\VECnab}{{\boldsymbol{\nabla}}}
\newcommand{\VECphi}{{\boldsymbol{\varphi}}}

\newcommand{\RR}{\mathbb{R}}

\newcommand{\Kout}{K_{\text{o}}}
\newcommand{\Kin}{K_{\text{i}}}

\newcommand{\hc}{h_{\text{c}}}
\newcommand{\ho}{h_{0}}
\newcommand{\hoff}{h_{\text{off}}}
\newcommand{\tildehc}{\tilde{h}_{\text{c}}}
\newcommand{\tildeho}{\tilde{h}_{0}}
\newcommand{\tildehoff}{\tilde{h}_{\text{off}}}

\newcommand{\alphai}{\alpha_{\text{i}}}
\newcommand{\alphao}{\alpha_{\text{o}}}
\newcommand{\alphac}{\alpha_{\text{c}}}
\newcommand{\alphat}{\alpha_{\text{t}}}

\newcommand{\deltai}{\delta_{\text{i}}}
\newcommand{\deltao}{\delta_{\text{o}}}

\begin{document}

\title{Balancing torques in membrane-mediated interactions: Exact results and
numerical illustrations}

\author{Martin Michael M\"uller}
\author{Markus Deserno}
\affiliation{Max-Planck-Institut f\"ur Polymerforschung, %
             Ackermannweg 10, %
             55128 Mainz, %
             Germany}
\author{Jemal Guven}
\affiliation{Instituto de Ciencias Nucleares, %
             Universidad Nacional Aut\'onoma de M\'exico, %
             Apdo.\ Postal 70-543, %
             04510 M\'exico D.F., %
             Mexico}

\date{\today}
\begin{abstract}
Torques on interfaces can be described by a divergence-free tensor
which is fully encoded in the geometry. This tensor consists of two
terms, one originating in the couple of the stress, the other
capturing an intrinsic contribution due to curvature. In analogy 
to the description of forces in terms of a stress tensor, 
the torque on a particle can be expressed as a line integral along a 
contour surrounding the particle. 
Interactions between particles mediated by a fluid membrane are studied
within this framework. In particular, torque balance places a strong
constraint on the shape of the membrane. Symmetric two-particle
configurations admit simple analytical expressions which are valid
in the fully nonlinear regime; in particular, the problem may be
solved exactly in the case of two membrane-bound parallel cylinders.
This apparently simple system provides some flavor of the remarkably
subtle nonlinear behavior associated with membrane-mediated interactions.
\end{abstract}

\pacs{87.16.Dg, 68.03.Cd, 02.40.Hw}

\maketitle


\bibliographystyle{plain}


\section{Introduction \label{sec:introduction}}

The crucial role of fluid membranes as a component of living cells
was recognized early on by biologists \cite{Lodish}. 
In recent years, the mechanics of membranes has become a fertile area 
of research among physicists as well \cite{HandbookofBioPhys}.
Many cellular tasks such as exo- or endocytosis \cite{Marsh01}, the 
formation of vesicles \cite{Robinson97,McMahonGallop05,Antonny06}, 
or the interaction with the cytoskeleton \cite{LuHi92} rely heavily 
on mechanical membrane properties and can be studied 
using the mathematical toolbox of theoretical physics. 
Various key questions are of an essentially geometrical nature:
what is the shape adopted by a membrane subjected to some specified
boundary conditions \cite{Seifert97}? How are forces and torques
transmitted through the membrane
\cite{surfacestresstensor,Guven04,Lomholt,Kozlov}? Which interactions does 
this imply 
\cite{mem_inter,geomsurf,Oettel,Kralchevsky,Koltover,gbp,inclusions,conicalinclusions,Weiklcyl,Fourspher,Kim,BisBis}?

To answer questions such as these, one approach is to apply an
effective formulation in which the membrane is modeled as a
two-dimensional surface in space. This approach is valid provided
the length scales of interest are much longer than the membrane
thickness which typically is a few nanometers. The energy of the
membrane is then completely determined by 
the geometry of the surface. In distinction to a simple fluid-fluid
interface where the energy is dominated by surface tension
\cite{deGennes}, bending elasticity now becomes important \cite{Canham,Helfrich}. 
It is possible, in principle, to determine equilibrium profiles by
solving a field equation (the ``shape equation'') which follows from
the minimization of the energy with respect to surface deformations.
The forces and torques transmitted by the membrane are then imprinted on
its shape.

The shape equation is a fourth order nonlinear partial differential
equation; as such, it can only be solved analytically in a few 
exceptional cases. 
One approach to sidestepping this obstacle is to expand the energy up 
to quadratic order. Unfortunately, this approach has the drawback that
it is only valid if the surface is a weak perturbation of some
underlying reference shape, such as a plane or a sphere. If the
membrane is highly curved, linearization becomes inadequate and
other strategies need to be developed.

In Ref.~\cite{surfacestresstensor} an alternative approach was
presented. It was shown that it is possible to relate the forces and
torques to the local membrane geometry without any need to solve the
shape equation itself. The link is formed by the surface stress tensor 
and its torque counterpart. This approach has already proven its value
when applied to the problem of surface-mediated interactions
\cite{mem_inter,geomsurf,Oettel}: 
particles bound to a membrane interact because they deform its shape
\cite{Kralchevsky,Koltover,gbp,inclusions,conicalinclusions,Weiklcyl,Fourspher,Kim,BisBis}. 
In the traditional approach to this problem based on energy, the
force between particles is calculated in three steps: first, the
equilibrium shape of the membrane is determined for a given
placement of the particles by solving the linearized shape equation.
The energy of the surface is then evaluated by integration, and
finally differentiated with respect to appropriate placement
variables of the bound particles. This approach relies heavily on
the validity of the linearization. 
Instead, one can also relate the force between the particles directly 
to the local geometry of the surface via the surface stress tensor.
It becomes possible to establish a host of valuable exact nonlinear
results \cite{mem_inter,geomsurf}.

In this context the particles were fixed by horizontal forces
allowing all other degrees of freedom, among them their vertical
positions or tilts, to equilibrate. To gain further insight a 
generalization is necessary which will be provided here: we will extend the
framework developed in Refs.~\cite{mem_inter} and \cite{geomsurf}
and explicitly address questions concerning the balance of torques.
This will impose strong additional constraints on the
shape of the membrane. To be concrete, we will examine symmetric
two-particle configurations, thereby obtaining a transparent
analytical expression describing the balance of torques which is
valid in the fully nonlinear regime. We will examine exactly the
interaction of two membrane-bound parallel cylinders. This is
possible because the membrane profile behaves like a one-dimensional
elasticum of Euler \cite{Eulerelastica}, a wormlike chain in the
language of polymer physics \cite{DNA1,DNA2,DNA3}; thus it can be
integrated exactly. This apparently simple system displays
remarkably subtle behavior. For strong membrane deformations 
the results of the small gradient approximation \cite{Weiklcyl} 
fail and nonlinear effects, such as multiple solutions for one given set of 
parameters, emerge. Moreover, the simple exponential decay of the force or 
the cylinder tilt angle as a function of separation, which is correct for 
small deformations \cite{Weiklcyl}, gives way to a more complicated dependency. 

The paper is organized as follows: in
Sec.~\ref{sec:torquesonsurfacepatches} we will discuss how torques
on surface patches can be determined within the framework of
Refs.~\cite{surfacestresstensor,Guven04}. To begin with, we will
treat a geometrical Hamiltonian which is given by a surface integral
over a local density. The behavior of this Hamiltonian under
boundary rotations will be used to identify the torque tensor. This
tensor will be written down explicitly for a fluid bilayer.
Surface-mediated interactions between particles on an asymptotically
flat membrane are considered in Sec.~\ref{sec:medinter}. Total force
and torque balance will be exploited to identify constraints on the
geometry of the membrane. In Sec.~\ref{sec:exactsolution} the
membrane geometry corresponding to two parallel cylinders will be
determined exactly and the balance of forces and torques examined in
detail. Finally, we summarize our main conclusions in
Sec.~\ref{sec:conclusions}.


\section{Torques on surface patches \label{sec:torquesonsurfacepatches}}

\subsection{Surface energetics \label{subsec:surfaceenergetics}}

Consider an interface 
which can be modeled as a two-dimensional surface in
three-dimensional Euclidean space. Such a surface $\Sigma$ is
described locally by its position $\VECX(\xi^1,\xi^2)\in\RR^3$,
where the $\xi^a$ are any suitable set of local coordinates on the
surface. The tangent vectors of $\Sigma$, $\VECe_{a}=\partial \VECX
/\partial \xi^{a}=\partial_{a}\VECX\;(a,b\in\{1,2\})$, form a local
coordinate frame; together with the extrinsic unit normal vector,
$\VECn=\VECe_{1}\times\VECe_{2}/|\VECe_{1}\times\VECe_{2}|$, they
define two geometrical tensors on the surface: the metric
$g_{ab}=\VECe_{a}\cdot\VECe_{b}$ and the extrinsic curvature
$K_{ab}=\VECe_{a}\cdot\partial_{b}\VECn$
\cite{DifferentialGeometry}. The trace of the extrinsic curvature,
$K=g_{ab}K^{ab}$, is the sum of the two principal curvatures whereas 
the Gaussian curvature $K_{\romG}=\det(K_{a}^{b})$ is their product. 
The covariant derivative on the surface will be denoted by $\nabla_{a}$,
the corresponding Laplacian by $\Delta=\nabla_{a}\nabla^{a}$.

We associate with the surface an energy which can be described as 
a surface integral over a local density $\mathcal{H}$ 
\begin{equation}
  H_{\Sigma}[\VECX] = \int_\Sigma \romd A \;
  \mathcal{H}(g_{ab},K_{ab}) \ ,
  \label{eq:surfacefunctional}
\end{equation}
where $\romd
A=\sqrt{\det{(g_{ab})}}\,\romd^{2}\xi=\sqrt{g}\,\romd^{2}\xi$ is the
infinitesimal area element. The density $\mathcal{H}$ depends
exclusively on surface scalars constructed using the metric and the
extrinsic curvature tensor.

To determine the equilibrium shape of the surface the
functional~(\ref{eq:surfacefunctional}) is minimized with respect to
deformations of $\Sigma$. These deformations are described by a
change in the embedding functions $\VECX\to\VECX+\delta\VECX$. The
corresponding variation of the Hamiltonian is
\cite{Guven04}
\begin{equation}
  \delta H_{\Sigma} = \int_{\Sigma} \romd A \; \mathcal{E} \, \VECn \cdot \delta\VECX +
    \int_{\Sigma} \romd A \; 
      \nabla_{a} \left[ \mathcal{H}^{ab} \VECe_b \cdot  \delta\VECn
        - \VECf^a \cdot \delta\VECX \right]
  .
  \label{eq:variationHamiltonian}
\end{equation}
Its bulk part is a surface integral over the Euler-Lagrange
derivative $\mathcal{E}(\mathcal{H})$ times the normal projection 
of the surface variation $\delta\VECX$. The second term is a surface
integral over a divergence and can thus be recast as a boundary
integral using the divergence theorem \cite{Frankel}. It involves
two parts: a part proportional to $\delta\VECX$ identifies the
surface stress tensor $\VECf^{a}$
\cite{Guven04,surfacestresstensor}; the second part proportional to
$\delta \VECn$ involves
$\mathcal{H}^{ab}=\partial\mathcal{H}/\partial K_{ab}$ and
originates from the partial integration of second derivatives of the
deformation. 
Note that additional terms would appear in the boundary term if
$\mathcal{H}$ also depended on \emph{derivatives} of $g_{ab}$ and
$K_{ab}$.

According to Noether's theorem, every continuous symmetry of the
Hamiltonian implies a conserved current. The one associated with
the translational invariance of $H$ is the surface stress tensor
$\VECf^{a}$. Its conservation describes the balance of forces across
curves drawn on the surface \cite{Guven04,surfacestresstensor}. In
an analogous way, the rotational invariance of $H$ implies the
existence of another conserved tensor $\VECm^{a}$ which, as will be
demonstrated in the next section, one can identify as the torque
tensor.


\subsection{Identification of the torque tensor
\label{subsec:identificationtorquetensor}}
\begin{figure}
\includegraphics[scale=0.7]{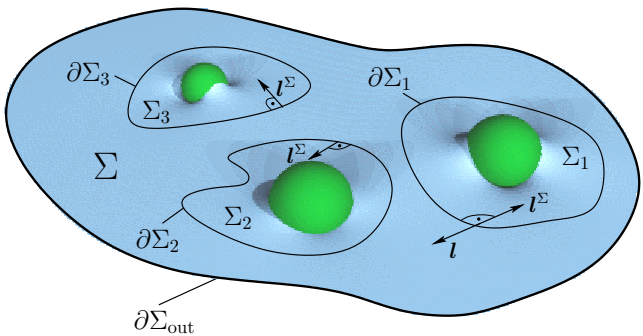}
  \caption{Surface region $\Sigma$ with 3 disjoint boundary components
  $\partial\Sigma_i$ and an outer limiting boundary
  $\partial\Sigma_{\text{out}}$.}\label{fig:identification}
\end{figure}

Consider a region of surface bounded by an outer limiting curve $\partial\Sigma_{\text{out}}$ 
and suppose that the bulk of this surface is in equilibrium with $N$ 
particles that are attached to it \cite{geomsurf}. 
We now choose a curve around each of the particles, labeling these 
curves $\partial\Sigma_{i}, i=1,\dots,N$. The outer curve $\partial\Sigma_{\text{out}}$ 
together with all $\partial\Sigma_{i}$ encloses the surface patch $\Sigma$ (see 
Fig.~\ref{fig:identification}). Considering this patch first allows us 
to use variation~(\ref{eq:variationHamiltonian}) to identify the torque tensor 
as no particles are attached to $\Sigma$ and every point on it is 
free to equilibrate. 

Under a constant infinitesimal rotation $\delta\VECalpha$ of one of
the boundaries $\partial\Sigma_i$, the position and normal vectors
change according to $\delta\VECX= \delta\VECalpha \times \VECX$ and
$\delta\VECn=\delta\VECalpha \times \VECn$. In this case, the boundary 
variation of the Hamiltonian~(\ref{eq:surfacefunctional}) is equal
to 
\begin{eqnarray}
  \delta H_{\Sigma}^{(i)} & \stackrel{(\ref{eq:variationHamiltonian})}{=} &
    -\delta\VECalpha \cdot \oint_{\partial\Sigma_{i}} \!\!\! \romd s \;
    l^{\Sigma}_{a} \left[ \VECX \times \VECf^{a}
      + \mathcal{H}^{ab} \VECe_{b} \times \VECn \right] \nonumber \\
  & = & -\delta\VECalpha \cdot \oint_{\partial\Sigma_{i}} \!\!\! \romd s \;
      l^{\Sigma}_{a} \VECm^{a}
  = -\delta\VECalpha \cdot \VECM_{\Sigma,\text{ext}}^{(i)}  \; ,
  \label{eq:variationconstantrotation}
\end{eqnarray}
where the surface integral was converted into a boundary integral on
the first line. The vector $\VECl^{\Sigma}=l^{\Sigma}_{a}\VECe^{a}$
is the unit vector which is normal to $\partial\Sigma_{i}$ and points out 
of the surface $\Sigma$; by construction it is tangential to $\Sigma$ (see
Fig.~\ref{fig:identification}). The variable $s$ measures the
arclength along $\partial\Sigma_{i}$.

In Eqn.~(\ref{eq:variationconstantrotation}), the boundary integral
is identified as the external torque $\VECM_{\Sigma,\text{ext}}^{(i)}$
acting on $\Sigma$ via its boundary $\partial\Sigma_{i}$: 
the corresponding change in energy is given by (minus) the scalar product of 
$\VECM_{\Sigma,\text{ext}}^{(i)}$ with the infinitesimal rotation angle 
$\delta\VECalpha$. 
The external torque $\VECM_{\text{ext}}^{(i)}$ \emph{on the surface patch} $\Sigma_{i}$ 
is then simply given by $-\VECM_{\Sigma,\text{ext}}^{(i)}$ due to torque balance. 
Notice that this argument does not require us to assume that the Euler-Lagrange 
equation is satisfied on $\Sigma_{i}$. 

To simplify notation we will restrict our considerations to patch $\partial\Sigma_{1}$ 
without loss of generality and write $\VECM_{\text{ext}}:=\VECM_{\text{ext}}^{(1)}$, with 
\begin{equation}
  \VECM_{\text{ext}} = \oint_{\partial\Sigma_{1}} \!\!\!  \romd s \; l_{a} \VECm^{a}
  \; ,
  \label{eq:torque}
\end{equation}
where $\VECl=l_{a}\VECe^{a}=-\VECl^{\Sigma}$. The surface tensor
\begin{equation}
  \VECm^{a} = \VECX \times \VECf^{a} +  \mathcal{H}^{ab} \VECe_{b} \times \VECn
  \;
  \label{eq:torquetensordef}
\end{equation}
is the covariantly conserved torque tensor. It consists of two
parts: a contribution due to the couple of the stress tensor
$\VECf^{a}$ about the origin as well as an intrinsic contribution
proportional to $\mathcal{H}^{ab}$ which originates only from curvature 
terms. 
The value of the former does depend on
the choice of origin; however, its intrinsic counterpart is
independent of this choice. One can easily check that the divergence
of $\VECm^{a}$, $\nabla_{a}\VECm^{a}$, is indeed zero in mechanical
equilibrium as required by consistency with Noether's theorem.

The occurence of curvature terms in the Hamiltonian is ultimately due 
to the fact that the physical surface is not infinitely thin but has an 
inhomogeneous force distribution along its transverse direction \cite{Lomholt}. 
If one captures this internal structure in the strictly two-dimensional 
Hamiltonian~(\ref{eq:surfacefunctional}), the curvature terms give rise to 
an intrinsic torque density and a non-vanishing normal component of the 
stress tensor. 

Note that the torque tensor, like the stress tensor, depends only on
geometric properties of the surface (modulo a contribution that is
due to a shift of the origin). It is thus always possible in
principle to determine the torques operating on a patch of surface
from a knowledge of the shape of the surface alone.  The snag is to
determine the correct shape.

It turns out, however, that exact geometric relations such as 
Eqns.~(\ref{eq:torque}) and (\ref{eq:torquetensordef}) are still
very useful, even when we lack the details of the shape, if
additional information, \eg a symmetry, is available. This was
shown in Refs.~\cite{mem_inter,geomsurf} in the context of
surface-mediated interactions between particles where the focus of
interest was the force. 
To see how this works for torques, let us first write down
Eqn.~(\ref{eq:torquetensordef}) for the surface models we are
interested in: soap films and fluid membranes.


\subsection{Special Cases \label{subsection:examples}}

\subsubsection{Soap film}
The Hamiltonian density of a soap film is given by $\mathcal{H} =
\sigma$, where $\sigma$ is the constant surface tension. This
implies that $\mathcal{H}^{ab}=0$. Thus the intrinsic torque
vanishes and, using  $\VECf^{a}=-\sigma \VECe^{a}$, one obtains
\cite{surfacestresstensor},
\begin{equation}
  \VECm^{a} = - \sigma \big( \VECX \times \VECe^{a} \big) \; .
  \label{eq:torquetensorsoapfilm}
\end{equation}
The torque on a surface patch $\Sigma_{1}$ is therefore given by
\begin{equation}
  \VECM_{\text{ext}} = - \sigma \oint_{\partial\Sigma_{1}} \!\!\! \romd s \;
    \big( \VECX \times \VECl \big) \; .
  \label{eq:exttorqueonsoapfilm}
\end{equation}
It is simple to interpret this expression: the tangential stress
$-\sigma\VECl$ provides a torque per unit length at every point of
the contour $\partial\Sigma_{1}$. The line integral along
$\partial\Sigma_{1}$ yields the total external torque on the surface
patch.


\subsubsection{Fluid lipid membrane}

A symmetric fluid membrane can be described by the Hamiltonian \cite{Canham,Helfrich}
\begin{equation}
  \mathcal{H} = \sigma + \frac{\kappa}{2} K^2 + \bar{\kappa} K_{\romG} \; ,
  \label{eq:Hamiltonianfluidmembrane}
\end{equation}
where $\sigma$ is, as before, the surface tension, $\kappa$ the bending rigidity, and 
$\bar{\kappa}$ the saddle-splay modulus. 
Together the two constants $\sigma$ and $\kappa$ provide a
characteristic length
\begin{equation}
  \lambda := \sqrt{\frac{\kappa}{\sigma}}
  \; ,
\end{equation}
separating short length scales over which bending energy dominates
from the large ones over which tension does.

The last term of the Hamiltonian density~(\ref{eq:Hamiltonianfluidmembrane}) can be written as 
the sum of a topological constant and a line integral over the geodesic curvature 
at the boundary of the membrane \cite{DifferentialGeometry}. 
It does not contribute to the membrane stress tensor and 
does not enter the shape equation. However, it gives a contribution to the 
torque tensor \cite{FournierMonge}: 
the derivative of the density~(\ref{eq:Hamiltonianfluidmembrane})
with respect to $K_{ab}$ is
\begin{equation}
  \mathcal{H}^{ab} = \kappa K g^{ab} + \bar{\kappa} (K g^{ab} - K^{ab}) 
  \; .
\end{equation}
Thus, the intrinsic torque does not vanish and, using the expression
for $\VECf^{a}$ from Ref.~\cite{geomsurf}, the torque tensor can be written as
\begin{eqnarray}
  \VECm^a & = &
    \Big[\kappa ( K^{ab} - \frac{1}{2}K g^{ab} )K
    - \sigma g^{ab}\Big] \big( \VECX \times \VECe_{b} \big) 
  \nonumber \\
  & & - \, \kappa (\nabla^a K) \big( \VECX \times \VECn \big)
  \nonumber \\
  & & + \, \Big[(\kappa + \bar{\kappa}) K g^{ab} - \bar{\kappa} K^{ab}\Big] 
    \big( \VECe_b \times \VECn \big) \; .
  \label{eq:torquetensormembrane}
\end{eqnarray}
Inserting this result into Eqn.~(\ref{eq:torque}) yields the external
torque
\begin{eqnarray}
  \VECM_{\text{ext}} & = &
  \oint_{\partial\Sigma_{1}} \!\!\! \romd s \;
    \Big\{ \Big[ \frac{\kappa}{2}(K_{\perp}^2 - K_{\|}^2) - \sigma \Big]
      \big( \VECX \times \VECl \big)\nonumber
  \\
  && \qquad \quad + \, \kappa K_{\perp\|}K \big( \VECX \times \VECt \big)
    - \, \kappa (\nabla_{\perp}K) \big( \VECX \times \VECn \big) \nonumber
  \\
  && \qquad \quad - \, \kappa K \VECt \Big\} \; ,
  \label{eq:exttorqueonmembrane}
\end{eqnarray}
on the membrane patch $\Sigma_{1}$. In
expression~(\ref{eq:exttorqueonmembrane}) the unit tangent vector
$\VECt=t^{a}\VECe_{a}=\VECn\times\VECl$ is introduced: it points along the
integration contour $\partial\Sigma_{1}$ and is perpendicular
to $\VECl$ and $\VECn$. The projections of the extrinsic curvature
onto the orthonormal basis of tangent vectors $\{\VECl,\VECt\}$ are
given by $K_{\perp}=l^{a}l^{b}K_{ab}, K_{\|}=t^{a}t^{b}K_{ab}$, and
$K_{\perp\|}=l^{a}t^{b}K_{ab}$. The symbol
$\nabla_{\perp}=l_{a}\nabla^{a}$ denotes the directional derivative 
along the vector $\VECl$.

The intrinsic part of the torque tensor adds a contribution to the 
integrand in Eqn.~(\ref{eq:exttorqueonmembrane}), which is proportional to 
the curvature $K$ and tangential to the contour of integration. Remarkably, the term 
$\VECM_{\text{ext}}^{\bar{\kappa}}=\int_{\partial\Sigma_{1}}\romd s\,l_{a}\VECm^{a}_{\bar{\kappa}}$ 
originating from the Gaussian curvature vanishes. The reason for this is that 
its integrand can be written as a derivative with respect to the arc-length $s$:
\begin{eqnarray}
  l_{a}\VECm^{a}_{\bar{\kappa}} & = &
  l_{a} \Big[\bar{\kappa} (K g^{ab} - K^{ab})\Big] 
    \big( \VECe_{b} \times \VECn \big) 
  = - \bar{\kappa} ( K_{\|} \VECt + K_{\perp\|}\VECl )
  \nonumber \\
  & = & - \bar{\kappa} ( t_{a} K^{ab} \VECe_{b}) 
  = - \bar{\kappa} \nabla_{\|} \VECn
  = - \bar{\kappa} \frac{\romd \VECn}{\romd s}
  \; ,
  \label{eq:Gaussiancurvaturetorquecontribution}
\end{eqnarray}
where the Weingarten equations $\nabla^{a}\VECn = K^{ab}\VECe_{b}$ were 
used and the symbol $\nabla_{\|}=t_{a}\nabla^{a}$ denotes the directional 
derivative along $\VECt$, \ie the derivative with respect to the 
arc-length. Integrating $\nabla_{\|} \VECn$ over the closed contour 
$\partial\Sigma_{1}$ yields zero and thus $\VECM_{\text{ext}}^{\bar{\kappa}}=0$, 
\emph{even if} an external torque is acting and \emph{even though} the 
torque tensor comprises terms proportional to $\bar{\kappa}$. 
This result implies that torques on membrane patches do not depend on the last term of the 
Hamiltonian density~(\ref{eq:Hamiltonianfluidmembrane}) involving the Gaussian curvature 
$K_{\romG}$. As shape and stresses are also independent of 
that term, we can neglect it in the following considerations. 

In addition, if $K$ is constant on $\partial\Sigma_{1}$, the intrinsic part of the torque tensor does not 
contribute to $\VECM_{\text{ext}}$ at all. This is because the integral of 
$\VECt=\romd \VECX / \romd s$ vanishes along any closed contour. 

It is straightforward to derive expressions analogous to
Eqn.~(\ref{eq:exttorqueonmembrane}) for any Hamiltonian of the
form~(\ref{eq:surfacefunctional}).

We will now show how the framework provided by the torque tensor may
be applied to the study of membrane-mediated interactions.


\section{Interface mediated interactions between particles on asymptotically
flat interfaces \label{sec:medinter}}

Particles bound to an interface may interact via the deformations
they impose on the surface geometry. Consider, for instance, a
multiparticle configuration such as the one depicted in
Fig.~\ref{fig:identification}: forces as well as torques act on each
of the particles due to the distortions in the shape caused by the
others. 
Such a situation can thus only be stationary if the particle
positions and orientations are constrained by external forces and
torques, respectively. As we have shown, these forces and torques
are determined completely by the surface geometry. The external
torque acting on the particle that adheres to the surface patch
$\Sigma_{1}$, for instance, 
is equal to the external torque $\VECM_{\text{ext}}$ acting on
the whole patch as given by Eqn.~(\ref{eq:torque}); this is because
the torque tensor is divergence-free on any part of the surface not
acted upon externally. The torque $\VECM^{(i)}$ we will consider in
the following is the torque on the particle $i$ mediated by the
interface \emph{counteracting} the corresponding external torque;
the former is evidently equal to minus the latter (for $i=1$ one
thus has $\VECM^{(1)}=-\VECM_{\text{ext}}$). Similar arguments apply
to the force $\VECF^{(i)}$ on the $i^{\text{th}}$ particle as
explained in detail in Refs.~\cite{mem_inter,geomsurf}.


\subsection{Balance of forces and torques \label{subsec:forcetorquebalance}}

The total force and the total torque must vanish if the
multiparticle configuration is in mechanical equilibrium:
\begin{equation}
  \sum^{N}_{i=1} \VECF^{(i)} = \VECF_{\text{out}} \qquad \text{and} \qquad
  \sum^{N}_{i=1} \VECM^{(i)} = \VECM_{\text{out}}
  \; ,
  \label{eq:totalforcetorquebalance}
\end{equation}
where $\VECF_{\text{out}}$ is the force and $\VECM_{\text{out}}$ the torque on
the outer boundary $\partial\Sigma_{\text{out}}$ (see again
Fig.~\ref{fig:identification}).

In the following, we will  consider, in particular, a symmetric
fluid membrane with nonvanishing surface tension $\sigma$ 
characterized by the Hamiltonian
density~(\ref{eq:Hamiltonianfluidmembrane}). Its ground state in the
absence of particles is an infinite flat plane with zero curvature. 
In this case the energy is proportional to the area and thus
infinite. This infinite constant in the energy will play no role in 
the sequel, and we will recalibrate the energy to set it to zero.

Particles that are bound to the membrane typically deform its shape
and increase its energy. These energy changes, however, must be
\emph{finite} whenever the forces and torques applied to fix the
particle configuration are finite. This implies that the interface
becomes asymptotically flat remote from the particles, even if
$\sigma$ is infinitesimally small (see
App.~\ref{app:scalingarguments}).

It is useful to decompose forces and torques into  a horizontal part
parallel to the asymptotic plane and a vertical part orthogonal to
it. When the surface tension vanishes, Kim \etal\ have shown that the
vertical force and the horizontal torque on the outer boundary at
infinity must vanish in equilibrium \cite{Kim}. In
App.~\ref{app:scalingarguments} we extend their discussion to treat
situations in which  $\sigma\ne 0$. The change is non-trivial: it
turns out that, whereas the vertical force vanishes as before, the
horizontal torque does not necessarily vanish. In antisymmetric
configurations, such as the ones that we will consider in the next
section, it is this horizontal contribution on the outer boundary
that makes it possible to balance an external torque.

In the following section, the complete set of equations describing
force and total torque balance [see
Eqns.~(\ref{eq:totalforcetorquebalance})] in a symmetrical
two-particle configuration will be derived.


\subsection{Two-particle configurations with symmetry
\label{subsec:twoparticleconfigurations}}

\begin{figure}
\includegraphics[scale=0.33]{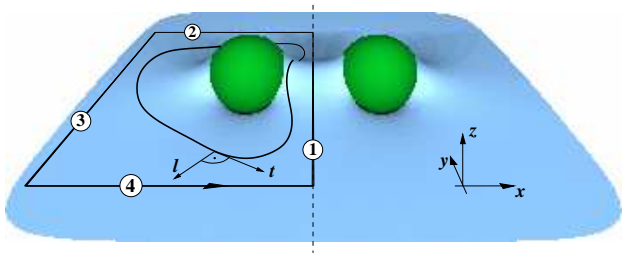}
  \caption{Two identical particles bound to an interface. As described
  in the text, it is possible to deform the contour of integration in order to
  exploit the available symmetries \cite{geomsurf}.}
\label{fig:twoparticles}
\end{figure}

Consider a symmetrical configuration of two identical particles
bound to an asymptotically flat surface as described in
Refs.~\cite{mem_inter,geomsurf} (see Fig.~\ref{fig:twoparticles}).
We label by $\{\VECx,\VECy,\VECz\}$ the Cartesian orthonormal basis
vectors of three-dimensional space $\RR^{3}$ adapted to the
asymptotic plane. The vectors $\VECx$ and $\VECy$ lie parallel to
this plane whereas $\VECz$ is its upward pointing unit normal. 

We will discuss two possible symmetries: mirror symmetry in the
$(y,z)$ plane (the symmetric case) or a twofold symmetry with
respect to the $y$ axis (the antisymmetric case). The line joining
corresponding points on the particles lies parallel to the $(x,z)$ plane. 
We place the origin of the coordinate system in the middle between the two 
particles on the intersection line of asymptotic and symmetry plane 
(symmetric case) or the line of symmetry (antisymmetric case), respectively.

External torques on the surface can either be vertical or horizontal. 
The former cannot be compensated by the outer boundary. The 
physical explanation for this is that the surface does not resist 
shear forces. Horizontal torques, however, can be balanced 
(see App.~\ref{app:scalingarguments}) which is necessary for a stable 
configuration: 
the separation between the particles is fixed by \emph{horizontal}
constraining forces. In the antisymmetric case, the two particles do
not generally lie on the same line of action. Thus the forces will
apply a horizontal torque $M_{y}$ to the surface which has to be balanced 
by the outer boundary. 
In the following, we will consider situations in which this torque is the only 
external torque \emph{on the entire surface}. 

This restriction does not exclude external vertical torques
$\VECM^{(i)}$ \emph{on the individual particles}; the symmetry will
not be broken as long as all these torques cancel. Think, for
instance, of a symmetric configuration consisting of two spheres on
a soap film with a saddle-shaped (quadrupolar) line of contact
\cite{Stamou_Fournier}. A vertical torque $M_{z}$ on one of the
particles is consistent with the symmetry so long as a torque
$-M_{z}$ is applied to its partner. 
Such possibilities may, in principle, be accommodated within the
formalism. Here, however, we will only consider situations in which
the particle orientations have equilibrated and these torques
vanish.

As explained previously, the torque on the left particle can be
obtained by integrating the appropriate projection of the torque
tensor along a contour surrounding the particle
\begin{equation}
  \VECM = -\oint_{1-4} \!\!\! \romd s \; l_{a}\VECm^{a}
  \; .
  \label{eq:torqueonleftparticle}
\end{equation}
We choose the same contour of integration 
as used in Refs.~\cite{mem_inter,geomsurf} (see
Fig.~\ref{fig:twoparticles}): it consists of four branches with
branch 1 lying on the symmetry plane/line and branches 2--4 pulled
open to infinity.

Using Eqn.~(\ref{eq:totalforcetorquebalance}) together with the
assumptions made above we conclude that
\begin{equation}
  \VECM=\VECX_{\text{p}}\times\VECF=M\VECy
  \; ,
  \label{eq:torquecondition}
\end{equation}
where $\VECF=F\VECx$ is the force on the left particle 
and $\VECX_{\text{p}}$ is the position vector pointing from the
fulcrum (\ie the point about which the torque is acting) to the
point where the force $\VECF$ is applied. We evaluate the torque
about the origin in the Euclidean coordinate system
$\{\VECx,\VECy,\VECz\}$. The torque on the outer boundary enclosing
the particles is then given by $\VECM_{\text{out}}=2\VECM$ in the
antisymmetric case; it vanishes in the symmetric case.

The condition~(\ref{eq:torquecondition}), together with the explicit
expression for the torque $\VECM$ derived here as well as its
counterpart for the force $\VECF$ derived in
Refs.~\cite{mem_inter,geomsurf}, permit one to establish nonlinear
relationships between geometrical quantities such as particle
penetration or height differences for a given model.


\subsection{Equilibrium conditions from torque balance \label{subsec:equilbriumcond}}

\subsubsection{General expressions for two finite-sized particles} 

For a fluid membrane, the torque $\VECM$ on a particle can be
obtained by inserting the corresponding torque
tensor~(\ref{eq:torquetensormembrane}) into
Eqn.~(\ref{eq:torqueonleftparticle}). As it evidently has to cancel
the external torque on the particle, it can also be read off
directly from the right hand side of
Eqn.~(\ref{eq:exttorqueonmembrane}) with the replacement of 
$\partial\Sigma_{1}$ by the contour $1-4$ and an additional minus 
sign out front.

The resulting expression can be simplified further if one takes into
account that the surface is asymptotically flat and possesses a
certain symmetry. 
For finite-sized particles it is shown in
App.~\ref{app:scalingarguments} that the curvature terms of the
torque tensor do not contribute to the torque on the membrane at
infinity; the only term in the expression for the torque due to
bending stems from branch 1. Furthermore, the term proportional to
$(\VECX\times\VECt)$ vanishes for the two symmetries because
$K_{\perp\|}\,K=0$ on branch 1 \cite{geomsurf}. The torque $\VECM$
on the left particle is thus given by
\begin{eqnarray}
  \VECM & = & \sigma 
    \oint_{1-4} \!\!\!\!
    \romd s \; (\VECX \times \VECl)
    - \, \kappa 
      \int_{1} \romd s \;
      \Big[ \frac{1}{2}(K_{\perp}^2 - K_{\|}^2) (\VECX \times \VECl)
  \nonumber \\
  && - \, (\nabla_{\perp}K) (\VECX \times \VECn) - K \VECt \Big] 
  \nonumber \\
  & = & \sigma 
    \oint_{2-4} \!\!\!\!
      \romd s \; (\VECX \times \VECl)
      - \kappa \int_{1} \romd s \; \Big[ \VECX \times \VECf - K \VECt \Big]
  \; . \;\;\;\;\;\;\;\;
  \label{eq:torqueonleftparticlemembrane}
\end{eqnarray}
where
$\VECf=[\frac{1}{2}(K_{\perp}^2 - K_{\|}^2)-\lambda^{-2}]\,\VECl-(\nabla_{\perp}K)\,\VECn$.

One now might naively expect that, in analogy to the curvature
terms, the term due to surface tension vanishes along branches 2--4:
the membrane becomes asymptotically flat at infinity and the
corresponding force has no vertical component. As $\VECX$ lies in
the asymptotic plane due to our choice of origin, it would appear to
be self-evident that the line integral vanishes.

If this argument were correct, it would also apply to the complete
outer boundary enclosing both particles and it would be impossible 
to balance the external torque. How then can a stable equilibrium be 
attained in the antisymmetric case if, as was argued earlier, an 
external torque is unavoidable?

Fortunately, a closer inspection of the integral shows that it does
not necessarily always vanish: 
for when branches 2--4 are sent to infinity, while the vertical
force component indeed does converge to zero, the length $|\VECX|$
of the corresponding lever arm goes to infinity simultaneously.
Additionally, one integrates along a contour whose length
depends linearly on $|\VECX|$. 
These two effects permit a finite value to remain: if we parametrize
the remote regions of the membrane in terms of a height function
above the asymptotic plane and consider its series
expansion~(\ref{eq:heightfunctionMongeparametrization}) for large
$|\VECX|$, a non-vanishing horizontal torque remains if the slowest
decaying term is of order $|\VECX|^{-1}$. In fact, from
Eqn.~(\ref{eq:heightfunctionMongeparametrization}) we obtain for the
height function 
\begin{eqnarray}
  h(|\VECX|,\varphi) & = & (C_{1}\cos{\varphi} + D_{1}\sin{\varphi}) \; |\VECX|^{-1}
  \nonumber \\
  && \; + \; \text{faster decaying terms}
  \; ,
  \label{eq:heightfunctionMongeparametrizationslowdecay}
\end{eqnarray}
where $C_{1}$ and $D_{1}$ are constants and $\varphi$ is the
azimuthal angle measured from the $x$ axis. Thus, branches 2--4 can 
contribute a torque if either of the two constants in 
Eqn.~(\ref{eq:heightfunctionMongeparametrizationslowdecay}) does not
vanish. A closer inspection of the line integral along branches 2--4
reveals that the term with coefficient $C_{1}$ contributes to a
torque about the $y$ axis whereas the one including $D_{1}$ does so
about $\VECx$ (\cf calculation in App.~\ref{app:scalingarguments}
for $\VECM_{\text{out}}$). The latter coefficient has to be equal to zero as 
we require the $x$ component of the torque $\VECM_{\text{out}}$ at the 
complete outer boundary to vanish (see again App.~\ref{app:scalingarguments}).

Bearing these considerations in mind, one obtains the following important relations
by combining Eqns.~(\ref{eq:torquecondition}) and (\ref{eq:torqueonleftparticlemembrane}).
\begin{subequations}
\begin{eqnarray}
  \VECx \cdot \VECM & = & 
    - \kappa \int_{1} \romd s \;
      \Big[ \VECX \cdot (\VECf \times \VECx) \Big]
  = 0
  \; ,
  \label{eq:Mxcondition}
  \\
  \VECy \cdot \VECM & = & 
    \sigma \oint_{2-4} \!\!\!
      \romd s \; \Big[ \VECX \cdot (\VECl \times \VECy) \Big]
  \label{eq:Mycondition} \\ 
  && - \kappa \int_{1} \romd s
      \Big[ \VECX \cdot (\VECf \times \VECy) - K (\VECy \cdot \VECt) \Big] 
    = F (\VECz \cdot \VECX_{\text{p}}) ,
  \nonumber \\
  \VECz \cdot \VECM & = &  \, -\kappa 
      \int_{1} \romd s \;
      \Big[ \VECX \cdot (\VECf \times \VECz) - K (\VECz \cdot \VECt) \Big]
  = 0
  \; , \;\;\;\;\;
  \label{eq:Mzcondition}
\end{eqnarray}
\end{subequations}
where $F=-\kappa\int_{1}\romd s\, (\VECf\cdot\VECx) -\sigma\int_{3}\romd s$
\cite{geomsurf}. These are the expressions referred to at the end of
the previous section. They place strong constraints on the geometry of the
membrane 
and can be simplified further for the two different symmetries.

\paragraph{Symmetric case}
In the symmetric case, the derivative of $K$ in the direction of
$\VECl$ along branch 1, $\nabla_{\perp}K$, is zero. The curvature tensor 
is diagonal in $(\VECl,\VECt)$ coordinates and thus $K_{\perp\|}$ 
vanishes. This implies that $K_{\|}\VECt$ is equal to 
$\nabla_{\|}\VECn$ [\cf Eqn.~(\ref{eq:Gaussiancurvaturetorquecontribution})], and 
$\int_{1}\romd s\, K_{\|}\VECt=\VECn|_{y=\infty}-\VECn|_{y=-\infty}=0$. 

Furthermore, 
branches 2--4 do not provide a torque: the term in the height
function~(\ref{eq:heightfunctionMongeparametrizationslowdecay}) that
could give rise to a torque about the $y$ axis is forbidden by the
symmetry, \ie its constant coefficient $C_{1}=0$. Thus, 
\begin{equation}
  \sigma 
    \oint_{2-4}
    \!\!\! \romd s \; (\VECX \times \VECl) = 0
  \; ,
\end{equation}
and Eqn~(\ref{eq:torqueonleftparticlemembrane}) simplifies to the expression 
\begin{equation}
  \VECM_{\text{sym}} = - \, \kappa 
    \int_{1} \romd s \;
    \Big\{ \Big[ \frac{1}{2}(K_{\perp}^2 - K_{\|}^2) -\lambda^{-2} \Big] 
    (\VECX \times \VECx) -K_{\perp}\VECt \Big\} 
  \; ,
\end{equation}
which depends only on geometrical properties of the membrane at the
symmetry plane. The force acting at each point along the mid-line is
horizontal and Eqn.~(\ref{eq:Mxcondition}) is fulfilled identically. 
The two relations~(\ref{eq:Mycondition}) and
(\ref{eq:Mzcondition}) turn into
\begin{eqnarray}
  - \, \kappa 
      \int_{1} \romd s \;
      \Big[ f_{\text{sym}} (\VECz \cdot \VECX) - K_{\perp} (\VECy\cdot\VECt) \Big]
    & = & F_{\text{sym}} (\VECz \cdot \VECX_{\text{p}}) \; , \;\;\;\;\;\;\;
  \label{eq:Myconditionsym}
  \\
  \, \kappa 
      \int_{1} \romd s \;
      \Big[ f_{\text{sym}} (\VECy \cdot \VECX) + K_{\perp} (\VECz\cdot\VECt) \Big]
    & = & 0
  \; ,
  \label{eq:Mzconditionsym}
\end{eqnarray}
with $f_{\text{sym}}=[\frac{1}{2}(K_{\perp}^2 - K_{\|}^2)-\lambda^{-2}]$ and
$F_{\text{sym}}=\sigma\Delta L-\frac{\kappa}{2}\int_{1}\romd s\,(K_{\perp}^2 - K_{\|}^2)$
where $\Delta L$ is the excess length of branch 1 compared to branch 3 \cite{geomsurf}.

\paragraph{Antisymmetric case}
In the antisymmetric case branch 1 is a straight line and
$K_{\perp}=K_{\|}=0$. The torque is now given by
\begin{equation}
  \VECM_{\text{antisym}} = \sigma 
    \oint_{1-4}
    \!\!\! \romd s \; (\VECX \times \VECl)
    + \kappa 
      \int_{1} \romd s \;
      \Big[ (\nabla_{\perp}K) (\VECX \times \VECn) \Big] \; .
\end{equation}
In contrast to the symmetric case, the torque does not possess an
internal part because the curvature $K$ vanishes everywhere along
the contour. 
The force acting at each point along the mid-line is perpendicular to the 
$y$ axis. As the origin lies on this line 
the only components of the torque transmitted through branch 1 which are
potentially non-vanishing are the $x$ and the $z$ component. We have chosen,
however, to set these components to zero [see previous Section and
Eqns.~(\ref{eq:Mxcondition},\ref{eq:Mzcondition})]. Thus, the contribution from branch 1
vanishes
\begin{equation}
 \sigma \int_{1} \romd s \; (\VECX \times \VECl)
   + \kappa \int_{1} \romd s \;
     \Big[ (\nabla_{\perp}K) (\VECX \times \VECn) \Big] = 0
 \; ,
 \label{eq:Mcontributionbranch1antisym}
\end{equation}
and the torque can be written as an integral along the boundaries
2-4
\begin{equation}
  \VECM_{\text{antisym}} = \sigma \int_{2-4}
    \!\!\! \romd s \; (\VECX \times \VECl)
  = F_{\text{antisym}} (\VECz \cdot \VECX_{\text{p}}) \VECy
  \; ,
  \label{eq:Myconditionantisym}
\end{equation}
where $F_{\text{antisym}}=\kappa\int_{1}\romd s\, [\lambda^{-2}(\VECx\cdot\VECl-1)
+ (\nabla_{\perp}K)(\VECx\cdot\VECn)]$ \cite{geomsurf}.

This equation implies a strong constraint  on the asymptotics of the
membrane. Far from the particles the term of leading order in the
height
function~(\ref{eq:heightfunctionMongeparametrizationslowdecay})
\emph{has} to be proportional to $|\VECX|^{-1}$. Otherwise, the
external torque would not be balanced. The coefficient $C_{1}$ has
to be finite, its value fixed by the external force responsible for
the torque.

If the height function of the membrane is known,
$\VECM_{\text{antisym}}$ can be easily calculated: 
in App.~\ref{app:scalingarguments} we have derived
expression~(\ref{eq:MoutMongeparametrization}) for
$\VECM_{\text{out}}$ in terms of the coefficients of the height
function~(\ref{eq:heightfunctionMongeparametrization})/
(\ref{eq:heightfunctionMongeparametrizationslowdecay}). From this
expression we obtain $\VECM_{\text{antisym}} =
(1/2)\VECM_{\text{out}} = \pi \sigma C_{1} \VECy$, thus fixing 
$C_{1}=M_{\text{out}}/(2\pi\sigma)$.


\subsubsection{Two infinitely-long cylinders}

\begin{figure*}
\includegraphics[scale=0.85]{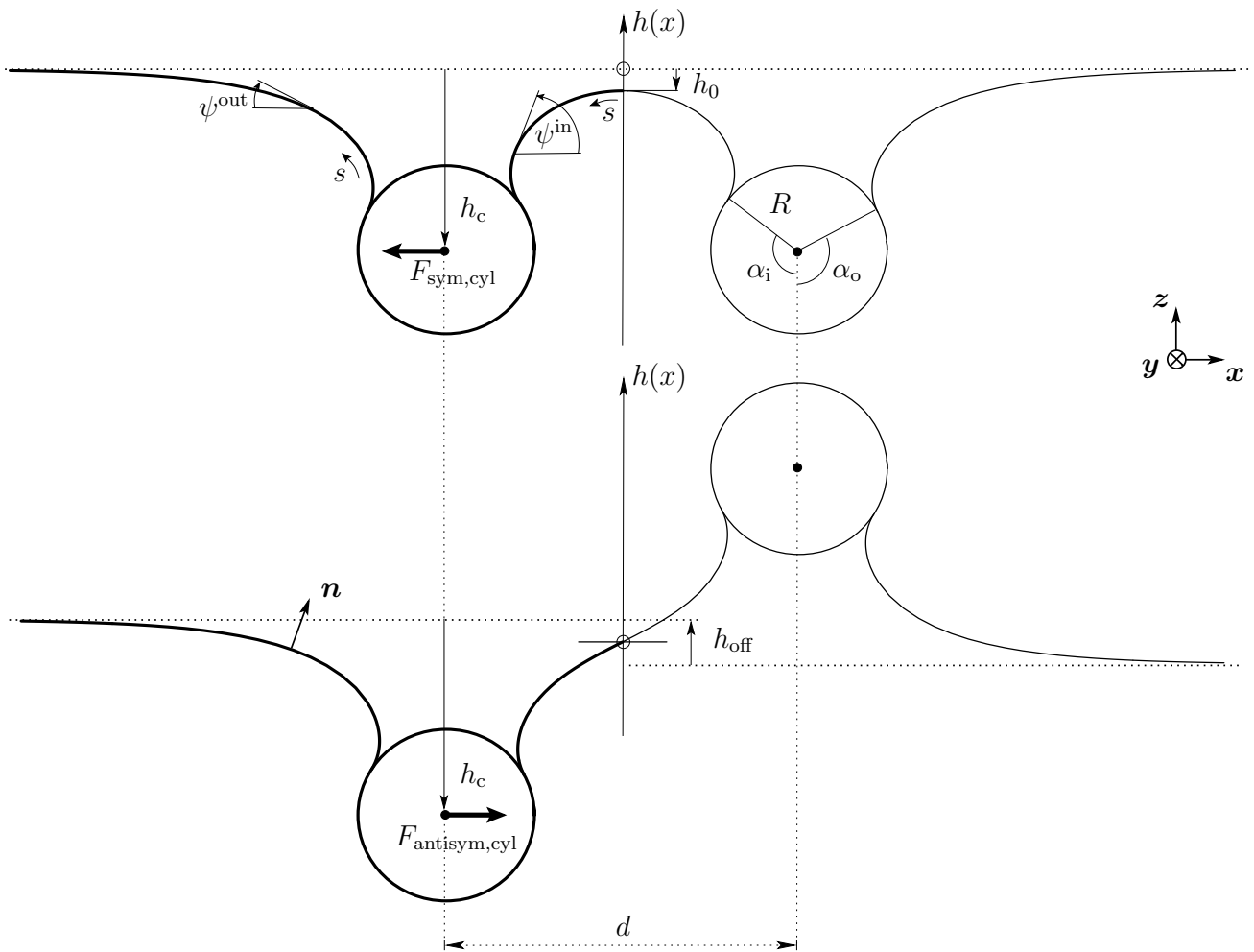}
\caption{Two parallel cylinders on a fluid membrane (symmetric and
antisymmetric case). The profiles are determined as explained in
Sec.~\ref{sec:exactsolution} using the following parameters:
$\alphac=\alphai+\alphao=240^{\circ}$, $d=4,\ R=1,\text{ and
}\lambda=1$. The point about which the torques are evaluated is
marked with $\circ$.} \label{fig:twocyltorque}
\end{figure*}

The situation changes if we consider two infinitely-long cylinders
of radius $R$ parallel to the $y$ axis that are separated by a 
distance $d$ (see Fig.~\ref{fig:twocyltorque}). 
The size of the particles is no longer finite. Thus, it is necessary
to exercise a little care when carrying over expressions derived for
finite-sized particles.

\paragraph{Symmetric case}

The symmetric case corresponds to two cylinders adhering to one side
of the membrane. The curvature parallel to branch 1, $K_{\|}$,
vanishes as branch 1 degenerates into a line. The contributions from
branch 2 and 4 cancel. 
No torque is exerted from branch 3: in contrast
to the configurations with two finite-sized particles, the shape
equation does not admit a term proportional to $|\VECX|^{-1}$ in the
height function (see Sec.~\ref{sec:exactsolution}). 

The torque per length $L$ of the cylinder with respect to the origin
of the Euclidean coordinate system $\{\VECx,\VECy,\VECz\}$ is thus
given by
\begin{equation}
  M_{\text{sym,cyl}}/L = \big( \frac{\kappa}{2} K_{\perp}^2
    - \sigma \big) \cdot \ho  + \kappa K_{\perp} \; ,
  \label{eq:MHelfrich}
\end{equation}
where the length $\ho$ is the distance between the origin and the
mid-point of the profile (see Fig.~\ref{fig:twocyltorque}). It is
positive if this point lies \emph{below} the origin.

Let $\hc:=-(\VECz\cdot\VECX_{\text{p}})$ be the vertical distance 
between the center of the cylinder and the asymptotic plane (see again
Fig.~\ref{fig:twocyltorque}). It is positive if the two cylinders
are located \emph{below} the asymptotic plane.

Torque balance establishes a relationship between $\ho$ and $\hc$,
involving only the geometry on the midline, $M_{\text{sym,cyl}}$
given by Eqn~(\ref{eq:MHelfrich}):
$M_{\text{sym,cyl}}
\stackrel{(\ref{eq:Myconditionsym})}{=} -\hc \, F_{\text{sym,cyl}}$
where $F_{\text{sym,cyl}}/L = - \frac{\kappa}{2}K_{\perp}^{2}$
\cite{geomsurf} implies
\begin{equation}
  \ho = \frac{\kappa K_{\perp} - \frac{\kappa}{2}K_{\perp}^{2} \hc}
    {\sigma - \frac{\kappa}{2} K_{\perp}^2}
  = \frac{K_{\perp} - \frac{1}{2} K_{\perp}^2 \hc}
    {\lambda^{-2} - \frac{1}{2} K_{\perp}^2}
  \; .
  \label{eq:ytorquecylmemsym}
\end{equation}

Eqn.~(\ref{eq:ytorquecylmemsym}) remains valid if the interface is a
soap film. The bending rigidity $\kappa$ then equals zero and we
immediately obtain the result that $\ho = 0$. This is confirmed by
the observation that the interface is flat everywhere and the
forces on the left and right cylinder vanish \cite{geomsurf}.

If we parametrize the profile as a height $h(x)$ above the
asymptotic reference plane, the curvature $K_{\perp}$ at branch 1 is
exactly equal to $-h''(0)$ in the symmetric geometry, where dashes
denote derivatives with respect to $x$. 
In Ref.~\cite{Weiklcyl} the height function $h(x)$ was determined at
the linearized level in the small gradient regime. Inserting it into
Eqn.~(\ref{eq:ytorquecylmemsym}) validates this relation
within the accuracy of the linear approximation (see
App.~\ref{app:linearcalc}).

\paragraph{Antisymmetric case}
In the antisymmetric case the two cylinders adhere to opposite sides of the membrane. 
The torque on the left cylinder is determined as follows: at branch
1, the line integral is zero. 
The contributions from branches 2 and 4 cancel as before. Thus, the
torque at branch 3 had better not vanish if the horizontal torque
due the force $F$ is to be balanced. 
However, the vertical force decreases faster than $1/x$ as
$x\to\pm\infty$ and thus cannot provide such a torque (\cf previous
paragraph and Sec.~\ref{sec:exactsolution}). How can torques balance
under these circumstances?

The solution to this apparent contradiction is the following: 
the asymptotic plane for positive values of $x$ does not necessarily
coincide with the one for negative values. This is because the
corresponding sections of the profile are \emph{disconnected}. 
It becomes possible for the membrane to shift vertically with an
offset $\hoff$ at the origin with respect to the asymptotic plane(s)
(see Fig.~\ref{fig:twocyltorque}). The torque is then simply given
by
\begin{equation}
  M_{\text{antisym,cyl}}/L =-\frac{\hoff}{2} \sigma
  \; .
\end{equation}
This offset may be related to $\hc$ as follows. One has 
$M_{\text{antisym,cyl}}\stackrel{(\ref{eq:Myconditionantisym})}{=}
-(\hc - \frac{\hoff}{2})F_{\text{antisym,cyl}}$ where
$F_{\text{antisym,cyl}}/L=\sqrt{\sigma^{2}+(\kappa\nabla_{\perp}K_{\perp})^{2}}-\sigma$
\cite{geomsurf}. One obtains
\begin{equation}
  \frac{\hoff}{2\hc} = 
    1 - \frac{1}{\sqrt{1+\lambda^{4}(\nabla_{\perp}K_{\perp})^{2}}} 
  \; ,
  \label{eq:ytorquecylmemantisym}
\end{equation}
where $\hc:=[-(\VECz\cdot\VECX_{\text{p}})+\hoff/2]$ is now measured 
from the \emph{left} asymptotic plane to the center of the cylinder 
(see again Fig.~\ref{fig:twocyltorque}).

Expanding the inverse square root in Eqn.~(\ref{eq:ytorquecylmemantisym}) 
up to zeroth and second order, respectively, yields a lower and an 
upper bound on the ratio of $\hoff$ to $\hc$:
\begin{equation}
  0 \le \frac{\hoff}{\hc} \le \lambda^{4} (\nabla_{\perp}K_{\perp})^{2}
  \; .
\end{equation}
Obviously, the offset $\hoff$ and $\hc$ have the same sign. If $\hc>0$, the left
asymptotic plane lies above the right one as depicted in Fig.~\ref{fig:twocyltorque};
for $\hc<0$ the situation is reversed.

By setting $\kappa=0$, we can again provide a check for consistency.
The offset is zero as in the symmetric case. This is in agreement
with the result from Ref.~\cite{geomsurf} that the interface is flat
and no force  acts on the cylinders.

In Ref.~\cite{Weiklcyl} the offset is set to zero in the ansatz for
the height function. On first inspection, this may appear to be an
error. However, it turns out to be consistent if
Equation~(\ref{eq:ytorquecylmemantisym}) is written in the small
gradient regime: the first non-vanishing term is of second order in
the smallness parameter. As the height function of \cite{Weiklcyl}
is itself correct only to first order, the two results agree at the
level of the approximation.

In this section several analytical conditions have been derived
which link different geometric properties of the interface profile
to each other. For the antisymmetric case we have seen that the
external torque is compensated either by a \emph{vertical force
component} (if the particles are finite) or by an \emph{offset} (if
the particles are infinitely-long cylinders). In the final part of
this paper we will demonstrate the value of the framework developed
in this section by showing how it can be applied to determine the
exact shape of the membrane with two adhering cylinders.


\section{Two cylinders on a fluid membrane -- exact solution
\label{sec:exactsolution}}


\subsection{Determining the profile\label{subsec:profilecalculation}}

\subsubsection{Shape equation}

If two parallel cylinders adhere to the membrane, the profile can be
decomposed into the following parts: two bound sections in which the
cylinder and membrane are in contact, an inner section between the
cylinders, and two outer sections that become flat for
$x\to\pm\infty$ (see Fig.~\ref{fig:twocyltorque}). The shape of the
bound parts is determined  by the geometry of the attached particle,
\ie a circular arc; the profiles of the free membrane sections are
determined by solving a nonlinear differential equation. 
Solving the equation itself is simple; the subtlety, as we will see,
is in the implementation of the boundary conditions. 

We will introduce the ``angle-arclength'' parametrization of the
profile: the angle $\psi(s)$ between the $x$ axis and the tangent to
the profile as a function of arclength $s$ completely describes the
shape of the membrane (see Fig.~\ref{fig:twocyltorque}). It is
connected to the curvature $K$ via the relationship $K = -
\dot{\psi}$, where the dot denotes a derivative with respect to $s$.

The nonlinear shape equation that determines the profile of the free membrane 
is given by \cite{Seifert97,planarelastica} 
\begin{equation}
  2 \ddot{K} + K^{3} - 2  \lambda^{-2} K = 0
  \; .
  \label{eq:shapeequation1D}
\end{equation}
It possesses the first integral
\begin{equation}
  \dot{K}^{2} + \frac{1}{4}K^{4} - \lambda^{-2} K^{2} = E
  \; ,
  \label{eq:1stintegralfictitiousparticle}
\end{equation}
where $E$ is a constant of integration.

\begin{figure*}
\includegraphics[scale=0.85]{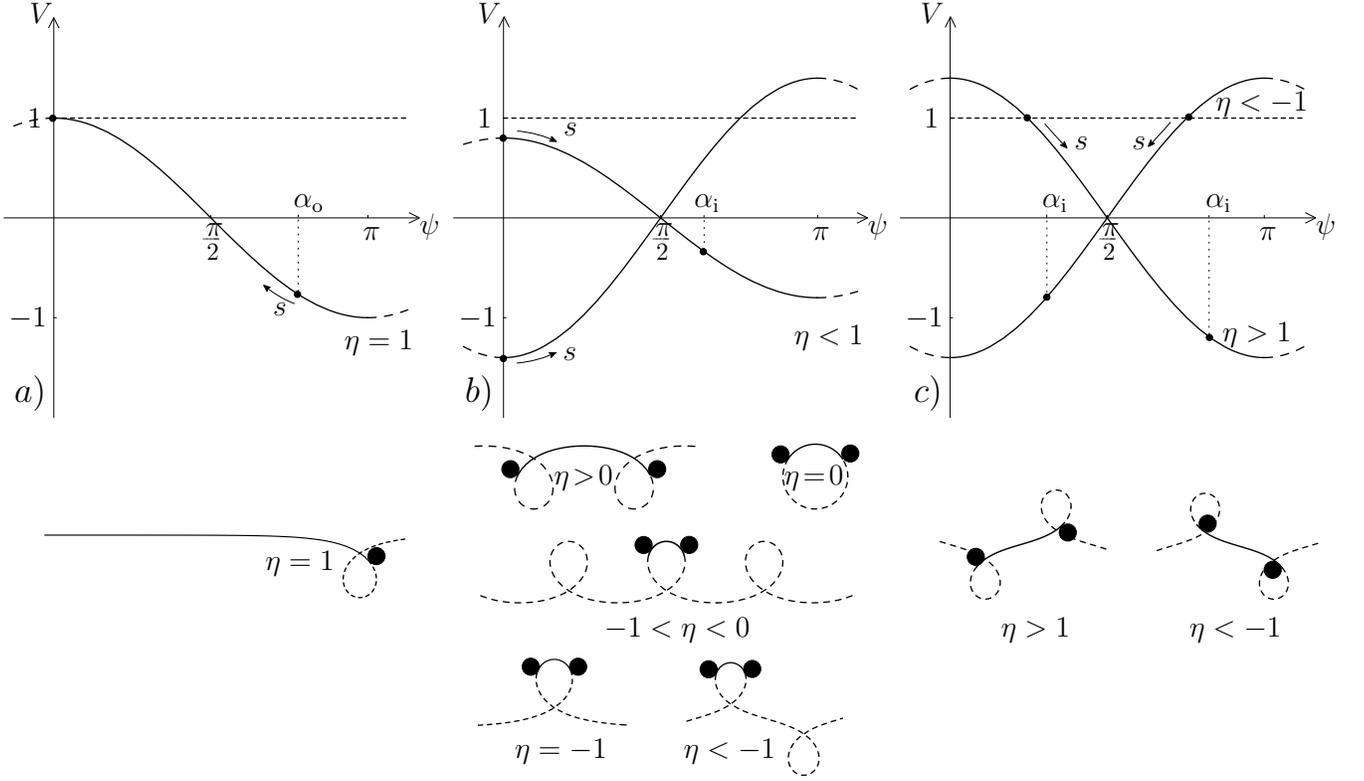}
\caption{\emph{Above:} Fictitious particle moving in the potential
$V(\psi)=\eta\cos{\psi}$ \;
a) outer section b) inner section - symmetric case c) inner section - antisymmetric
case; \emph{below:} corresponding
solutions~(\ref{eq:psioutersection},\ref{eq:psiinnersectionsym},\ref{eq:psiinnersectionantisym}) of
Eqn.~(\ref{eq:sineGordon1stintegral}).}
\label{fig:potential}
\end{figure*}

It is straightforward to integrate
Eqn.~(\ref{eq:1stintegralfictitiousparticle}) after a separation of
variables to determine $s$ as a function of $K$ which can be
inverted to express $K$ as a function of $s$. An additional
integration yields the shape of the membrane, $\psi(s)$.
One can do better however by taking advantage of the fact that
the force $\VECF_{\text{mem}}$ per length $L$ of the cylinder at
every point of the membrane is not only constant but also horizontal
on each membrane section. Thus,
\begin{equation}
  \VECF_{\text{mem}}/L = -l_{a}\VECf^{a} = -\Big(\frac{1}{2}\kappa K^{2} - \sigma \Big)\,\VECl
  +\kappa \dot{K}\,\VECn  = f_{\text{mem}} \VECx
  \; .
  \label{eq:forceonmembranepatch}
\end{equation}
Projecting this equation onto $\VECn$ yields, with
$\eta=f_{\text{mem}}/\sigma=\text{const.}$,
\begin{equation}
  \lambda^{2}\ddot{\psi} - \eta \sin{\psi} = 0
  \; .
  \label{eq:sineGordon}
\end{equation}
where we have used $\VECx\cdot\VECn=-\sin{\psi}$ and $K=-\dot{\psi}$.
Eqn.~(\ref{eq:sineGordon}) 
appears in a number of other physical applications which include the
motion of a simple pendulum \cite{LandauLifschitz1}, the shape of
a fluid meniscus under gravity \cite{shapeeqn_meniscus}, 
or the behavior
of elastic rods (Euler elastica) \cite{Eulerelastica}. Despite
appearances, it is completely equivalent to
Eqn.~(\ref{eq:1stintegralfictitiousparticle}): both are of the same
order in derivatives of $\psi$; furthermore, as we will show, the
constant of integration $E$ can be written in terms of the scaled
force $\eta$ and vice versa. Eqn.~(\ref{eq:sineGordon}), however,
has the advantage that it possesses a first integral with an
integration constant that can be deduced from
Eqn.~(\ref{eq:forceonmembranepatch}): the projection of the latter
onto $\VECl$ yields, with $\VECx\cdot\VECl=\cos{\psi}$,
\begin{equation}
  \frac{\lambda^{2}}{2}\dot{\psi}^{2}+\eta\cos{\psi} = 1
  \; .
  \label{eq:sineGordon1stintegral}
\end{equation}
Eqn.~(\ref{eq:sineGordon1stintegral}) can be interpreted as an
energy balance of a fictitious particle moving in the potential
$V(\psi)=\eta\cos{\psi}$, with displacement variable $\psi$, mass $\lambda^{2}$ 
and energy 1. The details of the ``motion" depend on the value of $\eta$
(see below and Fig.~\ref{fig:potential}). From
Eqn.~(\ref{eq:sineGordon1stintegral}) follows
\begin{equation}
  K = \pm \lambda^{-1}\sqrt{2(1-\eta\cos{\psi})}
  \; .
  \label{eq:curvatureKasfunctionofpsi}
\end{equation}
It is simple to check that $K$ satisfies the shape
equation~(\ref{eq:shapeequation1D}). The constants $\eta$ and $E$
are not independent; the relation between them,
$\eta=\pm\sqrt{E\lambda^{4}+1}$, is found by substituting $K$ into
Eqn.~(\ref{eq:1stintegralfictitiousparticle}).

In the following, we will determine the solutions of
Eqn.~(\ref{eq:sineGordon1stintegral}) for the free sections of the
profile if the cylinders are separated by a distance $d$. Due to the
symmetry it is sufficient to consider only the left half of the
membrane (where $x<0$, see Fig.~\ref{fig:twocyltorque}). While the
outer sections are qualitatively identical for the two symmetries,
the inner sections differ. Eqn.~(\ref{eq:sineGordon1stintegral}) is
a first order differential equation involving one unknown constant
$\eta$. Thus two boundary conditions are required for each section.


\subsubsection{Outer section}
The boundary conditions for the outer section are the following: the
free profile leaves the cylinder at a fixed contact angle
$\psi(0)=\alphao$. We assume without loss of generality that $\alphao>0$ and
$\dot{\psi}\le0$ \cite{negativeangles}. At infinity, both $\psi$ and
$\dot{\psi}$ vanish. If we insert these two conditions into
Eqn.~(\ref{eq:sineGordon1stintegral}), we obtain $\eta=1$ on this
section. This implies that $E=0$ everywhere along the profile. The
corresponding potential $V(\psi)$ is plotted in
Fig.~\ref{fig:potential}\ a): the solution with $\psi(0)= \alphao$
decreases monotonically to $\psi=0$ (where $V=1$) as $s\to\infty$.

The corresponding shape equation can be solved in terms of
elementary functions. One obtains
\begin{equation}
  \psi^{\text{out}}(\tilde{s})
    = 4 \arctan{\Big[ \tan{\frac{\alphao}{4}} \, \text{e}^{-\tilde{s}} \Big]}
  \; ,
  \label{eq:psioutersection}
\end{equation}
where $\tilde{s}:=s/\lambda$. 
Incidentally, the function~(\ref{eq:psioutersection}) also describes
the shape of a straight fluid meniscus that approaches a solid
surface at the angle $\alphao$ \cite{deGennes,shapeeqn_meniscus}.

The function~(\ref{eq:psioutersection}) decays exponentially with
increasing arclength. Thus, in contrast to the case of finite-sized
particles, the horizontal torque due to a vertical force component,
$|M|\approx \sigma |x| \sin{\psi^{\text{out}}} \approx \sigma |x|
\psi^{\text{out}}$ vanishes at infinity (see
App.~\ref{app:scalingarguments} and
Sec.~\ref{subsec:equilbriumcond}).


\subsubsection{Inner section}

We first establish the connection between the total arc-length of
the inner profile $2 s_{\text{mid}}$ and the contact angle between
the cylinder and the inner membrane,  $\alphai$. If arc-length is
measured from the mid-line, we have
\begin{equation}
  \frac{\tilde{d}}{2} - \tilde{R} \sin{\alphai}
    = \int_{0}^{\tilde{s}_\text{mid}} \!\!\!\romd \tilde{s} \; \cos{\psi}(\tilde{s})
  \; ,
  \label{eq:conditionsmid}
\end{equation}
where all lengths, as before, are scaled with $\lambda$. 
This condition determines the value of $\tilde{s}_{\text{mid}}$
implicitly in terms of $\alphai$.

We again consider only positive values of $\psi$. The solution for
negative angles are obtained by reversing the sign of the functions
(\ref{eq:psiinnersectionsym}) and (\ref{eq:psiinnersectionantisym})
which corresponds to a reflection of the profile in the $(x,y)$
plane. Note also that $K=+\dot{\psi}$ for all shapes of the inner
section due to the chosen orientation of the arclength.


\paragraph{Symmetric case}

In the symmetric case, $\psi(0)$ is equal to zero. From
Eqn.~(\ref{eq:sineGordon1stintegral}) we thus obtain $\eta\le 1$.
For $\eta=1$, the cylinders are infinitely far apart and do not
interact. 
If we omit this trivial case, the following solutions for 
different values of $\eta$ are obtained \cite{Eulerelastica,DNA3,etalower1sym} (see
Fig.~\ref{fig:potential}\ b):
\begin{equation}
  \psi^{\text{in}}_{\text{sym}}(\tilde{s}) =
  \left\{ \begin{array}{ll}
  \pi + 2
    \operatorname{am}{\Big( \tilde{s} \sqrt{\frac{\eta}{m}} - \operatorname{K}(m) , m \Big)}
  & \hspace*{-3ex},\ 0<\eta<1
  \\
  \sqrt{2} \tilde{s} & ,\ \eta=0
  \\
  2 \operatorname{am}{\Big( \tilde{s} \sqrt{\frac{|\eta|}{m}} , m \Big)}
  & \hspace*{-5ex},\ -1<\eta<0
  \\
  4 \arctan{\Big[\tanh{(\frac{\tilde{s}}{2}})\Big]} & ,\ \eta=-1
  \\
  \arccos{\Big[ 1- \frac{2}{m} \,
    \operatorname{sn}^{2}{\Big( \tilde{s} \sqrt{|\eta|}, \frac{1}{m} \Big)} \Big]}
  & ,\ \eta<-1
  \end{array} \right.
  \; ,
  \label{eq:psiinnersectionsym}
\end{equation}
where $\operatorname{am}(s,m)$ is the Jacobi amplitude with parameter $m$ and
$\operatorname{sn}(s,m)=\sin{[\operatorname{am}(s,m)]}$. The symbol $K(m)$ denotes
the complete elliptic integral of the first kind \cite{Abramowitz}. The parameter $m$
is given by $m:=\frac{2|\eta|}{1+|\eta|}\in[0,2[$.


\paragraph{Antisymmetric case}

If the two cylinders adhere antisymmetrically, $\dot{\psi}$ must
vanish at the mid-line. Thus,
\begin{equation}
  \cos{\psi_{\text{mid}}} \stackrel{(\ref{eq:sineGordon1stintegral})}{=} \frac{1}{\eta}
  \; ,
\end{equation}
where $\psi_{\text{mid}} = \psi(0)$ is the angle at the mid-line.
The case where $\psi_{\text{mid}}$ equals zero or $180^{\circ}$
corresponds again to the trivial solution with the two cylinders
infinitely far apart. No solution exists for
$\psi_{\text{mid}}=90^{\circ}$ as the scaled force $\eta$ has to
remain finite in an equilibrium situation. If
$0<\psi_{\text{mid}}<90^{\circ}$, $\eta>1$ and, for 
$-\frac{1}{\sqrt{\eta}}\operatorname{K}(\frac{1}{m})\le\tilde{s}
\le\frac{1}{\sqrt{\eta}}\operatorname{K}(\frac{1}{m})$ \cite{Eulerelastica,DNA3}
(see Fig.~\ref{fig:potential}\ c)
\begin{subequations}
  \label{eq:psiinnersectionantisym}
  \begin{equation}
  \psi^{\text{in}}_{\text{antisym}}(\tilde{s}) =
  \arccos{\Big\{ \frac{2}{m} \,
    \operatorname{sn}^{2}{\Big[ \tilde{s} \sqrt{\eta}
    - \operatorname{K}(\frac{1}{m}) , \frac{1}{m} \Big]} -1 \Big\}}
  \; ;
  \label{eq:psiinnersectionantisym1}
  \end{equation}
if it is greater than $90^{\circ}$ and lower than $180^{\circ}$, $\eta<-1$ and, for $-\frac{1}{\sqrt{|\eta|}}\operatorname{K}(\frac{1}{m})\le\tilde{s}
\le\frac{1}{\sqrt{|\eta|}}\operatorname{K}(\frac{1}{m})$\cite{Eulerelastica}
(see again Fig.~\ref{fig:potential}\ c)
  \begin{equation}
   \psi^{\text{in}}_{\text{antisym}}(\tilde{s}) =
   \arccos{\Big\{ 1 - \frac{2}{m} \,
    \operatorname{sn}^{2}{\Big[ \tilde{s} \sqrt{|\eta|}
    - \operatorname{K}(\frac{1}{m}) , \frac{1}{m} \Big]} \Big\}}
  \; .
  \label{eq:psiinnersectionantisym2}
  \end{equation}
\end{subequations}
The parameter $m$ is defined as above and varies now between 1 and 2.


\subsubsection{Boundary conditions at the cylinders}

To obtain the profile of the membrane for given separation $\tilde{d}$
and cylinder radius $\tilde{R}$, 
one has to determine the values of the scaled force $\eta$ and the
contact angles $\alphao$ and $\alphai$. The value of $\eta$ for any
given $\alphai$ is determined implicitly by the requirement that
\begin{equation}
  \psi^{\text{in}}(\tilde{s}_{\text{mid}})=\alphai
  \; , 
  \label{eq:conditionalphai}
\end{equation}
where $\tilde{s}_{\text{mid}}$ is itself implicitly given by
Condition~(\ref{eq:conditionsmid}); the values of $\alphai$ and
$\alphao$ depend on the boundary conditions at the cylinder. 
We will discuss two possibilities here: we either fix the area of
contact between the cylinders and the membrane, or consider
attachment due to a finite adhesion energy $w$ per area between the
membrane and the cylinders.

\paragraph{Fixed area of contact}
Suppose that the area of contact is fixed to 
\begin{equation}
  \alphac=\alphao+\alphai=\text{const}
  \; . 
  \label{eq:conditionfixedcontactarea}
\end{equation}
Torque balance will fix the relative values of $\alphao$ and $\alphai$.

The torque about the cylinder axis has to vanish in equilibrium.
This is the case if the total energy $E_{\text{tot}}(\alphai)$ of
the system exhibits a local extremum. The corresponding torque
balance can be written as
\begin{eqnarray}
  0 & = & \tilde{\Kin} - \tilde{\Kout} - \tilde{R} (\eta \cos{\alphai} - \cos{\alphao})
  \nonumber \\
  & \stackrel{(\ref{eq:curvatureKasfunctionofpsi})}{=} &
    \tilde{\Kin} - \tilde{\Kout} +\frac{\tilde{R}}{2} (\tilde{\Kin}^{2} - \tilde{\Kout}^{2})
  \; ,
  \label{eq:torquebalanceatcylinder}
\end{eqnarray}
where $\tilde{\Kin}:=\lambda\Kin$ and $\tilde{\Kout}:=\lambda\Kout$.
Eqn.~(\ref{eq:torquebalanceatcylinder}) has two solutions:
$\tilde{\Kin}=\tilde{\Kout}$ and
$\tilde{\Kin}=-\tilde{\Kout}-2/\tilde{R}$. The latter implies that
either $\tilde{\Kin}$ or $\tilde{\Kout}$ is smaller than
$-1/\tilde{R}$. Since the cylinder has a curvature of $-1/\tilde{R}$
in scaled units and membrane and cylinder must not intersect, the
second solution can be ruled out. The two contact curvatures must
agree:
\begin{equation}
  \tilde{\Kin}=\tilde{\Kout}
    =-\frac{\romd \psi}{\romd \tilde{s}}^{\text{out}}\Big|_{\tilde{s}=0}
    \stackrel{(\ref{eq:psioutersection})}{=} 2\sin{(\alphao/2)}
  \; ,
  \label{eq:Kout}
\end{equation}
and
\begin{equation}
  \eta \cos{\alphai} \stackrel{(\ref{eq:torquebalanceatcylinder})}{=} \cos{\alphao}
  \label{eq:etafromtorquebalance}
  \; .
\end{equation}
The values of $\eta$, $\alphai$, $\alphao$, and $\tilde{s}_{\text{mid}}$ 
for given separation and cylinder radius can be determined numerically by solving the 
conditions~(\ref{eq:conditionsmid}), (\ref{eq:conditionalphai}), 
(\ref{eq:conditionfixedcontactarea}), and (\ref{eq:Kout}) simultaneously. 
As an example, the two profiles
for $\tilde{d}=4$, $\tilde{R}=1$, and $\alphac=240^{\circ}$
are plotted in Fig.~\ref{fig:twocyltorque} .

\paragraph{Adhesion balance}
If the cylinders attach to the membrane due to a finite adhesion
energy $w$, the contact curvature condition holds (see \cite[Sec.\ 12, problem
6]{LaLi_elast}, \cite{Seifert90}, and \cite{boundarycondition}):
\begin{equation}
  \tilde{\Kin}=\tilde{\Kout}=-1/\tilde{R}+\sqrt{\tilde{w}}
  \; ,
  \label{eq:adhesionbalance}
\end{equation}
where $\tilde{w}:=\frac{2w\lambda^{2}}{\kappa}$. The curvatures are equal as in 
case $a$ and torque balance~(\ref{eq:torquebalanceatcylinder}) is fulfilled 
automatically.
The contact angle $\alphao$ can be determined from Eqn.~(\ref{eq:Kout}).
Finally, $\eta$, $\alphai$, and $\tilde{s}_{\text{mid}}$ can be calculated using  
Eqns.~(\ref{eq:conditionsmid}), (\ref{eq:conditionalphai}), and 
(\ref{eq:adhesionbalance}).

The adhesion energy $\tilde{w}$ and the wrapping angle $\alphac$ are
conjugate variables. Setting one of them to a constant value implies
that the other will adjust in equilibrium. Questions of stability
depend on which of the two variables is fixed: a profile found to be
stable under constant $\alphac$ is not necessarily stable under the
constant $\tilde{w}$. To avoid problems of this kind we will focus, in
the following, on the constant wrapping angle scenario.

Note that the boundary condition (\ref{eq:Kout}) 
does not always fix a unique profile. The total energy
$E_{\text{tot}}(\alphai)$ of the system may possess more than one minimum; thus
profiles which are locally stable may exist. In the next paragraph
we will see that this is indeed the case for certain ranges of
values of $\tilde{d}$, $\tilde{R}$, and $\alphac$.


\subsection{Conditions from torque balance\label{subsec:condfromtorquebalance}}
Now that we possess the complete profile we
can go back and reexamine Eqns.~(\ref{eq:ytorquecylmemsym}) and
(\ref{eq:ytorquecylmemantisym}) and study the behavior of
$\tildeho:=\ho/\lambda$ and $\tildehoff:=\hoff/\lambda$
as a function of distance $\tilde{d}$ for different values of $\tilde{R}$ 
and $\alphac$. 

\setcounter{paragraph}{0}
\paragraph{Symmetric case}
\begin{figure}
\includegraphics[scale=0.4]{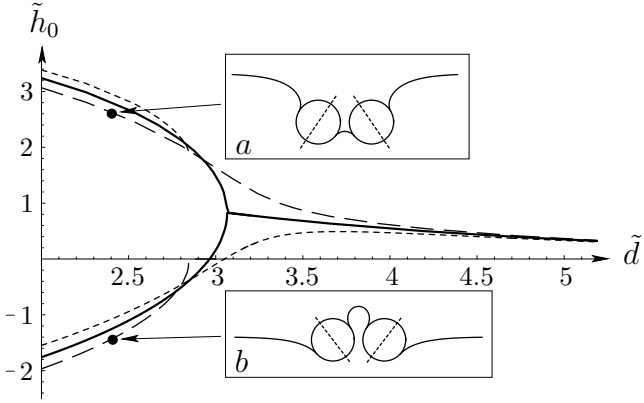}
\caption{The height $\tildeho$ at the mid-line as a function of separation $\tilde{d}$
  for $\tilde{R}=1$ and $\alphac=170^{\circ}$ (long dashes), $180^{\circ}$ (bold solid line),
  and $190^{\circ}$ (short dashes). At $\tilde{d}=2$ the two cylinders are in
  contact with each other and cannot draw nearer. 
  For $\tilde{d}\lesssim 3$ two locally stable solutions
  can be found which correspond to different angles $\alphai$. In the insets $a$ and $b$
  the two solutions for $\tilde{d}=2.4$ and $\alphac=170^{\circ}$ are plotted where
  $\alphai=50.9^{\circ}$ ($a$) and $122.2^{\circ}$ ($b$), respectively. Profile $a$
  corresponds to the global energy minimum here.}
\label{fig:hozoom}
\end{figure}
In the symmetric case, $\tilde{K}_{\perp}=\sqrt{2(1-\eta)}$ and
Eqn.~(\ref{eq:ytorquecylmemsym}) can be written as
\begin{equation}
  \tildeho = \eta^{-1}[\sqrt{2(1-\eta)} - (1-\eta) \tildehc]
  \; ,
\end{equation}
where
\begin{equation}
  \tildehc = 2\sin{(\alphao / 2)}-\tilde{R}\cos{\alphao}
  \; .
  \label{eq:hc}
\end{equation}
The height $\tildeho$ can be determined as a function of separation 
$\tilde{d}$ for fixed cylinder radius $\tilde{R}$ and different wrapping 
angles $\alphac$. 
If the value of $\alphac$ is small, the solution
coincides with the result of the small gradient
approximation~(\ref{eq:hosmallgradientapproximation}) as expected.
The value of $\tildeho$ falls off exponentially with the decay
length $\ell:=2\lambda$ for increasing separation $\tilde{d}$.

For higher angles $\alphac$, however, the behavior changes at small
values of  $\tilde{d}$ due to the breakdown of the small gradient
approximation (see Fig.~\ref{fig:hozoom} for $\tilde{R}=1$): 
the energy $E_{\text{tot}}(\alphai)$ 
exhibits two minima instead of one \cite{maximum}. These minima
correspond to distinct stable profiles with different $\alphai$ and
$\tildeho$. If $\alphac=180^{\circ}$, they possess the same energy.
If $\alphac<180^{\circ}$, the global minimum is located at the
smaller value of $\alphai$ and the corresponding $\tildeho$ is
always positive, \ie the mid-point lies below the reference plane
(see again Fig.~\ref{fig:hozoom}). For larger contact angles
$\alphac$ this behavior is reversed: the higher value of $\alphai$
and the lower value of $\tildeho$ correspond to the global energy
minimum. The mid-point may now even lie above the reference plane. 

\begin{figure}
\includegraphics[scale=0.45]{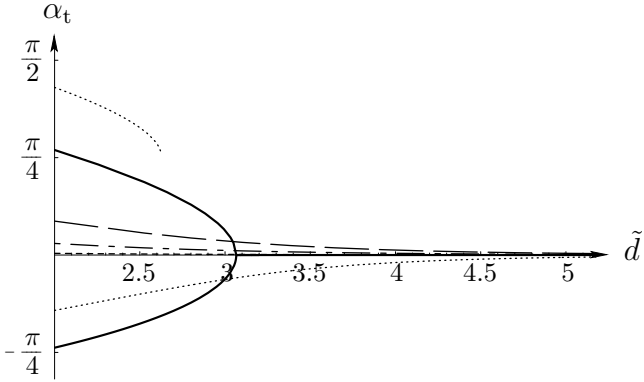}
\caption{The tilt angle $\alphat$ (symmetric case) as a function of 
separation $\tilde{d}$ for $\tilde{R}=1$ 
and $\alphac=10^{\circ}$ (short dashes), $45^{\circ}$ (dashed-dotted line),
$90^{\circ}$ (long dashes), $180^{\circ}$ (bold solid line),
and $240^{\circ}$ (dotted line). 
}
\label{fig:alphatiltsym}
\end{figure}

In Fig.~\ref{fig:alphatiltsym} the tilt angle $\alphat:=\frac{\alphac}{2}-\alphai$
is plotted as a function of distance $\tilde{d}$ for $\tilde{R}=1$.
It decays exponentially with length $\lambda$ 
at large separations. If $\alphac<180^{\circ}$ the tilt angle of the
global energy minimum is always positive which corresponds to
profiles such as the one plotted in Fig.~\ref{fig:hozoom}, $a$. If
$\alphac>180^{\circ}$, the tilt angle of the global energy minimum
is negative. For $\alphac=180^{\circ}$ and for \emph{all} $\tilde{d}\gtrsim 3$,
$\alphat$ is equal to zero; at $\tilde{d}\approx 3$ the function
bifurcates into two branches of same energy. A closer inspection of
Eqn.~(\ref{eq:etafromtorquebalance}) explains these findings: either
$\alphai$ is exactly $90^{\circ}$ and thus $\alphat=0$, or
$\eta=-1$. At the bifurcation point both conditions hold. For
$\tilde{d}\lesssim3$ the profiles with $\eta=-1$ are local minima of
same energy whereas $\alphat=0$ corresponds to the intermediate
maximum we do not discuss further here.


\paragraph{Antisymmetric case}

\begin{figure}
\includegraphics[scale=0.6]{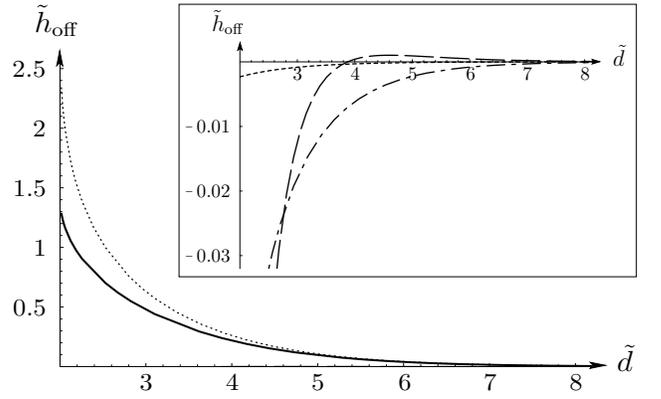}
\caption{The offset $\tildehoff$ as a function of separation $\tilde{d}$ for
$\tilde{R}=1$ and different contact angles. The line styles are chosen as in
Fig.~\ref{fig:alphatiltsym}.}
\label{fig:hoff}
\end{figure}
From Eqns.~(\ref{eq:ytorquecylmemantisym}) and (\ref{eq:sineGordon})
one obtains with $\nabla_{\perp}K_{\perp}=-\ddot{\psi}|_{\tilde{s}=0}$, $K_{\perp}=0$,
and $\eta>1$ 
\begin{equation}
  \frac{\tildehoff}{2 \tildehc} = 1 - \frac{1}{\eta} 
  \; ,
\end{equation}
where $\tildehc$ is given by Eqn.~(\ref{eq:hc}).

If the two cylinders are closer than $2\tilde{R}$, one additionally
has to check for every stable profile whether the particles overlap
or not. To avoid complications of this kind we will only consider
separations $\tilde{d}\ge 2\tilde{R}$. In contrast to the symmetric
case, one then finds a single solution for given $\tilde{d}$,
$\tilde{R}$, and $\alphac$.

In Fig.~\ref{fig:hoff} $\tildehoff$ is plotted as a function of
distance $\tilde{d}$. For small contact angles, $\tildehoff$ is
negative which implies that the asymptotic plane on the left lies
below the one on the right and below the center of the left
cylinder. For intermediate values of $\alphac$ (such as $90^{\circ}$
if $\tilde{R}=1$) $\tildehoff$ can be either positive or negative
depending on the separation of the cylinders (see again
Fig.~\ref{fig:hoff}). If $\alphac$ is further increased,
$\tildehoff$ is positive for all separations and the profiles
resemble the one which is plotted in the lower part of
Fig.~\ref{fig:twocyltorque}.

The tilt angle $\alphat$ is always equal to
zero for $\alphac=180^{\circ}$. The solution $\eta=-1$ which also solves
Eqn.~(\ref{eq:etafromtorquebalance}) is now forbidden due to the symmetry.
For $\alphac>180^{\circ}$ the tilt angle is positive; for $\alphac<180^{\circ}$
it is negative. In both cases it decays exponentially with length $\lambda$.


\subsection{Forces between the cylinders\label{subsec:forceoncylinders}}

\begin{figure}
\includegraphics[scale=0.43]{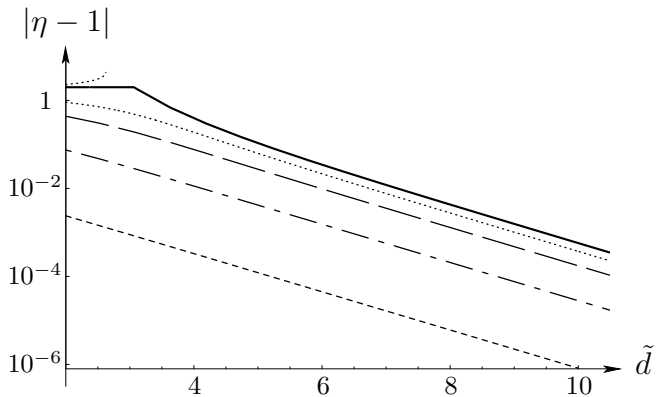}
\caption{Absolute value of the scaled force $|\eta - 1|$ as a function of separation
$\tilde{d}$ for $\tilde{R}=1$ and different contact angles (symmetric case). The line
styles are again chosen as in Fig.~\ref{fig:alphatiltsym}.}
\label{fig:forcesym}
\end{figure}
%
\begin{figure}
\includegraphics[scale=0.4]{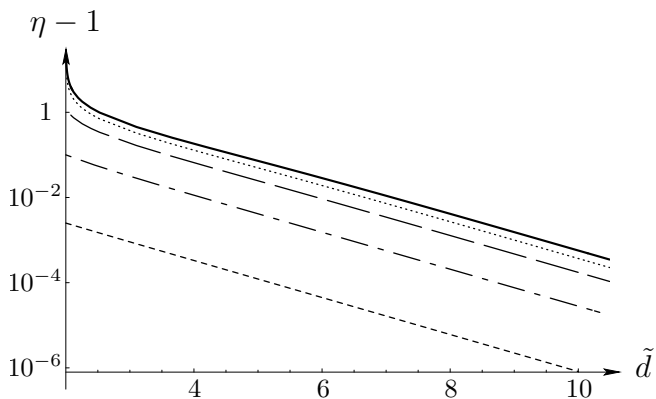}
\caption{Scaled force $\eta - 1$ as a function of separation
$\tilde{d}$ for $\tilde{R}=1$ and different contact angles (antisymmetric case). The line
styles are chosen as in Fig.~\ref{fig:alphatiltsym}.}
\label{fig:forceanti}
\end{figure}

Using the stress tensor \cite{geomsurf,mem_inter}, the force on the
left cylinder is given by
\begin{equation}
  F_{\text{cyl}}/L \stackrel{(\ref{eq:forceonmembranepatch})}{=} f_{\text{mem}} -\sigma
    =  \sigma (\eta - 1)
  \; .
\end{equation}
Thus, one has only to determine the value of $\eta$; the force follows directly.

\begin{figure}
\includegraphics[scale=0.45]{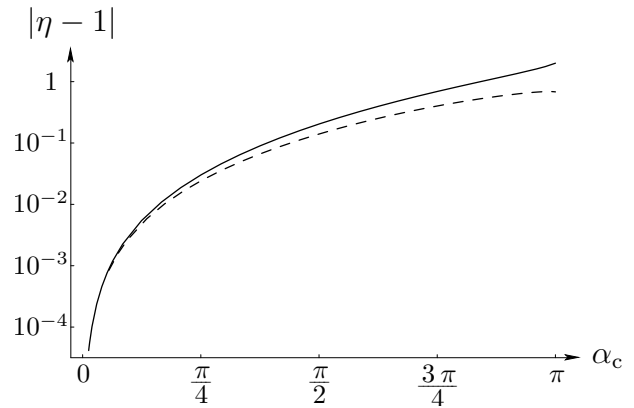}
\caption{$|\eta -1|$ as a function of $\alphac$ for $\tilde{d}=3$
and $\tilde{R}=1$ in the symmetric case. The small gradient
approximation is plotted with a dashed line, the numerically exact
solution with a solid line.} \label{fig:Weiklcomp}
\end{figure}

In Figs.~\ref{fig:forcesym} and \ref{fig:forceanti} the absolute value of the scaled force,
$|\eta-1|$,
is plotted for the symmetric and antisymmetric case. At 
large separations it decays exponentially as predicted by the
linearized theory \cite{Weiklcyl}. The underlying small gradient
approximation, however, breaks down if the curvatures along the
profile become too high, \ie for large contact area and small
separation. In Fig.~\ref{fig:Weiklcomp} the exact solutions are
compared with the small gradient approximation for the symmetric
geometry by plotting $|\eta-1|$ as a function of $\alphac$ for
$\tilde{d}=3$ and $\tilde{R}=1$. The approximate solution
increasingly underestimates the force as the contact angle is
increased.

As soon as the small gradient approximation breaks down, different
features appear: in the antisymmetric case, $|\eta-1|$ increases
faster for decreasing separation than a small gradient
approximation would anticipate. In the symmetric case, two solutions
can be found (\cf previous paragraph). The global energy minimum
typically corresponds to the solution with the smaller absolute
value of the force. Remarkably, the force is constant
[$F_{\text{cyl}}/(L\sigma)=-2$] for $\alphac=180^{\circ}$ and
$\tilde{d}\lesssim 3$.


\section{Conclusions\label{sec:conclusions}}

We have extended the formalism which was originally developed in 
Refs.~\cite{Guven04,surfacestresstensor} and primarily applied to forces
\cite{mem_inter,geomsurf} to study torques on surface patches.
The underlying torque tensor depends only on
geometric properties of the surface (modulo the contribution that is
due to a shift of the origin). Therefore, it is possible to relate
torques on particles which are bound to an interface directly to the
interface geometry. For the case of the Helfrich Hamiltonian we have 
explicitly worked out the torque tensor. Remarkably, its contribution 
from the Gaussian curvature part, even though nonzero, vanishes upon 
closed-loop integration and therefore does not transmit torques between 
particles. 

By exploiting torque and force balance in the context of
membrane-mediated interactions, several analytical conditions on the
interface geometry have been derived for two-particle configurations
obeying a certain symmetry. In the symmetric two-particle case all
external torques on the whole surface can be set to zero; in
contrast to that, such an external torque is inevitably applied 
in the antisymmetric case due to the forces which fix the separation between the
particles. 
Total torque balance determines the asymptotic behavior of the
membrane for both configurations. Particularly in the antisymmetric
case, the torque is either compensated by a vertical force component
(if the particles are finite) or by an offset between the two
asymptotic planes (if the particles are infinitely-long cylinders).
Complications stemming from additional external torques (\eg due to
a magnetic field if the two particles are dipoles) have not been
considered here explicitly but can be included with ease into the
relevant expressions.

To see how the rather abstract expressions from the first part can
be applied, the two-cylinder case ("1D problem") has subsequently been solved in
the nonlinear regime. The expressions obtained from torque balance
could be used to calculate the values of the offset (antisymmetric
case) and the height at the mid-plane (symmetric case).
Finally, the force on the cylinders could be identified within the
formalism as being essentially one of the parameters that had
to be determined in the calculation of the profile.

The simple 1D problem already displays remarkably subtle behavior. For 
large deformations the small gradient approximation fails 
which is due to the intrinsic nonlinearity of the problem. The sign of the 
force, however, is predicted correctly by the approximate solution. 
For two spheres on the membrane (2D problem), this 
need not be the case: although linear calculations predict 
a repulsion for the symmetric configuration \cite{conicalinclusions}, the 
nonlinear expression for the force $F_{\text{sym}}$ does not rule out that 
the interaction between the particles is attractive for higher deformations 
[cf. expression for $F_{\text{sym}}$ below Eqn.~(\ref{eq:Mzconditionsym})].  
While it is difficult to determine the sign analytically, 
coarsed-grained computer simulations suggest that a two-body 
attraction emerges if the deformations are high enough 
\cite{membranesimulations}.


\begin{acknowledgments}

The authors would like to thank IPAM at UCLA where part of this work
was discussed. MMM is also grateful for the hospitality during his
stay in Mexico City. Special thanks go to Ben Reynwar and Stefan
Bohlius for helpful discussions. Partial support to JG from CONACyT
grant 51111 as well as DGAPA PAPIIT grant IN119206-3 is
acknowledged.

\end{acknowledgments}


\appendix

\section{Forces and torques on the outer boundary \label{app:scalingarguments}}

Consider a symmetric fluid membrane with a finite number of finite-sized
particles bound to it. What is the asymptotic behavior of profile properties 
such as the curvature $K$ and its derivatives far away from the particles?

To answer this question let us adapt a line of argument originally introduced
in \cite{Kim} for the case of vanishing surface tension.
The excess membrane energy $E$ induced by the particles is finite.
From this requirement
it immediately follows that $K$ has to be square integrable and thus
vanishes far away from the particles. This implies that the behavior 
of the membrane is determined completely by surface tension there. 
Thus, the membrane becomes asymptotically flat. This fact allows us to use
linearized Monge gauge to describe its behavior remote from the
particles: the membrane profile is described in terms
of its height $h(r,\varphi)$ above the asymptotic plane where $r$
and $\varphi$ are the cylindrical coordinates defined on that plane; 
we choose to place the origin in the middle of the surface region to which the particles adhere.
For large $r$, the curvature $K$ is then equal to the negative (base
plane) Laplacian of $h$, $K=-\VECnab^{2}h$.

The energy can be written as $E = E_{\sigma} + E_{\kappa} + E_{\bar{\kappa}}$, where
$E_{\sigma}=(\sigma/2) \int \romd\varphi \, \romd r \, r \,(\VECnab h)^{2}$
and
$E_{\kappa}=(\kappa/2) \int \romd\varphi \, \romd r \, r \,(\VECnab^{2} h)^{2}$. 
The contribution due to the Gaussian curvature, $E_{\bar{\kappa}}$, is finite as it 
can be written as a topological constant plus a line integral over the geodesic 
curvature at the membrane boundary \cite{DifferentialGeometry}. 
The energies $E_{\sigma}$ and $E_{\kappa}$ are both positive and thus finite as well. 
The corresponding membrane shape equation, expressed in terms of $K$, is given 
by $\VECnab^{2} K = \lambda^{-2} K$ which has the general solution \cite{conicalinclusions}
\begin{eqnarray}
  K(r,\varphi) & = & a_{0} K_{0}(r/\lambda)
  \nonumber \\
  && \,  + \, \sum_{n=1}^{\infty} (a_{n}\cos{n\varphi} + b_{n}\sin{n\varphi}) \, K_{n}(r/\lambda)
  \nonumber \\
  && \, + \, \sum_{n=1}^{\infty} (c_{n}\cos{n\varphi} + d_{n}\sin{n\varphi}) \, I_{n}(r/\lambda)
  \; , \;\;\;\;\;\;
\end{eqnarray}
where $I_{n}$ and $K_{n}$ are modified Bessel functions of the first
and second kind, respectively \cite{Abramowitz}. The functions
$I_{n}$ tend to infinity as $r\to\infty$. Thus, $c_{n}$ and $d_{n}$
must vanish. The asymptotic expansion of $K_{n}$ for large $r$ is
given by
\begin{equation}
  K_{n}(r) = \sqrt{\frac{\pi}{2r}}\text{e}^{-r} \Big( 1+ \frac{4n^2 -1}{8r} + \ldots \Big)
  \; ,
\end{equation}
and shows that the functions $K_{n}(r)$ decay essentially 
exponentially for increasing $r$.

The force $\VECF_{\text{out}}$ on the outer boundary is a line
integral over the appropriate projection of the stress tensor,
$\int_{\partial\Sigma_{\text{out}}}\romd s \, l_{a}\VECf^{a}$. Since
$l_{a}\VECf^{a}$ depends on derivatives and powers of $K$ plus a
surface tension term \cite{geomsurf}, the contribution to the force
due to curvature vanishes as $r\to\infty$. The only surviving
contribution at infinity is a horizontal pulling force proportional
to $\sigma$.

Although the vertical force vanishes at infinity, an external torque
$\VECM_{\text{out}}$ may still act on the outer membrane boundary.
The reason for this is that the length of the lever arm increases
linearly with $R$. 
The origin of the torque is the surface tension and it is given by
\begin{eqnarray}
  \VECM_{\text{out}} & = & \sigma \int_{0}^{2\pi} \!\!\! \romd \varphi \;
    R [R\VECr + h(R,\varphi) \VECz] \times \VECl 
  \nonumber \\
  & = & \sigma \int_{0}^{2\pi} \!\!\! \romd \varphi \;
    R \{-h(R,\varphi)l_{\varphi} \VECr + [h(R,\varphi)l_{r} - R l_{z}] \VECphi
  \nonumber \\
  && \qquad\qquad  +  R l_{\varphi} \VECz\}
  \; ,
  \label{eq:MoutpreMongeparametrization}
\end{eqnarray}
where $\VECl=l_{r}\VECr+l_{\varphi}\VECphi+l_{z}\VECz$ in
cylindrical coordinates with basis vectors
$\{\VECr,\VECphi,\VECz\}$. The unit vectors $\VECr$ and $\VECphi$
lie on the reference plane whereas $\VECz$ is normal to it. The
moving trihedron of the boundary curve consists of the vectors
$\VECl$, $\VECt$, and $\VECn$ which are perpendicular to each other.
In particular, $\VECl$ may be expanded in terms of the height
function $h$ and its derivatives:
\begin{equation}
  \VECl = \frac{\VECr
     - \frac{1}{qR}\big[(\partial_{\varphi}h) (\partial_{r}h)\big]\big|_{r=R} \VECphi
     + \frac{1}{q}(\partial_{r}h)\big|_{r=R} \VECz}
    {\sqrt{1
    + \Big\{\frac{1}{qR}\big[(\partial_{\varphi}h) (\partial_{r}h)\big]\big|_{r=R} \Big\}^{2}
    + \Big\{\frac{1}{q}(\partial_{r}h)|_{r=R} \Big\}^{2}}}
  \; ,
  \label{eq:VEClMongeparametrization}
\end{equation}
where $q:=1+\Big[\frac{(\partial_{\varphi}h)|_{r=R}}{R}\Big]^{2}$.

The height function $h(r,\varphi)$ satisfies the linearized shape
equation $\VECnab^{2}(\VECnab^{2} h) = \lambda^{-2} \VECnab^{2} h$.
The solution of this equation consistent with a finite value for
$E_{\sigma}$ is \cite{conicalinclusions}
\begin{eqnarray}
  h(r,\varphi) & = & C_{0} + A_{0}K_{0}(r/\lambda)
  \nonumber \\
  && \,  + \, \sum_{n=1}^{\infty} (A_{n}\cos{n\varphi} + B_{n}\sin{n\varphi}) \, K_{n}(r/\lambda)
  \nonumber \\
  && \,  + \, \sum_{n=1}^{\infty} (C_{n}\cos{n\varphi} + D_{n}\sin{n\varphi}) \, r^{-n}
  \; .
  \label{eq:heightfunctionMongeparametrization}
\end{eqnarray}
Inserting Eqn.~(\ref{eq:VEClMongeparametrization})
and (\ref{eq:heightfunctionMongeparametrization}) into the
equation of the torque~(\ref{eq:MoutpreMongeparametrization}) one obtains
\begin{equation}
  \VECM_{\text{out}} = \sigma \int_{0}^{2\pi} \!\!\! \romd \varphi \; R
  \Big[\frac{2(C_{1}\cos{\varphi} +  D_{1}\sin{\varphi})}{R} \VECphi
    + \mathcal{O}\Big(\frac{1}{R^{2}}\Big) \Big]
  \; .
  \label{eq:Moutpre2Mongeparametrization}
\end{equation}
The unit vector $\VECphi$ still depends on the coordinate $\varphi$.
To evaluate the integral over $\varphi$ in
(\ref{eq:Moutpre2Mongeparametrization}), we rewrite $\VECphi$ as
$\VECphi=-\sin{\varphi}\,\VECx + \cos{\varphi}\,\VECy$, where
$\VECx$ and $\VECy$ are the Cartesian coordinates parallel to the
asymptotic plane. 
We finally obtain for $R\to\infty$
\begin{equation}
  \VECM_{\text{out}} = 2 \pi \sigma (-D_{1} \VECx + C_{1} \VECy)
  \; .
  \label{eq:MoutMongeparametrization}
\end{equation}
Hence, the horizontal torque is in general not equal to zero
even though the vertical force is. Its contributions stem from those terms
of the height function~(\ref{eq:heightfunctionMongeparametrization}) which are
proportional to $r^{-1}$ and decay slowest for increasing distance from the particles. 
Conversly, the existence of an external torque forces the surface to display a very 
slow $1/r$ decay.


\section{Calculations on the linearized level \label{app:linearcalc}}

In this appendix we will show that Eqn.~(\ref{eq:ytorquecylmemsym}) is consistent 
with the results of Ref.~\cite{Weiklcyl}. To this end we calculate $h(0)=-\ho$ 
and compare it with the right hand side of (\ref{eq:ytorquecylmemsym}).

Due to the symmetry it is sufficient to consider only the left
cylinder. Its axis is located at $x=-d/2$ (see
Fig.~\ref{fig:twocyltorque}). For $-d/2-\deltao\le x \le
-d/2+\deltai$, cylinder and membrane are in contact 
($\delta_{\text{i,o}}=R\sin{\alpha_{\text{i,o}}}$). In small
gradient approximation the shape of the outer membrane segment
($x\le d/2-\deltao$) is given by \cite{Weiklcyl}
\begin{equation}
  h_{\text{out}}(x) = B \exp{(-|x + d/2|/\lambda)}
  \; ,
\end{equation}
where $B=-\deltao\lambda / R$ to first order in $\deltao$ (strictly speaking, 
the expansion is in $\deltao/\lambda$ here and below). The
membrane segment adhering to the left cylinder has the circular
profile
\begin{equation}
  h_{\text{cyl}}(x) = -\Big[\hc + \sqrt{R^{2} - (x + d/2)^{2}}\Big]
  \; ,
\end{equation}
where $\hc$ is defined  in Fig.~\ref{fig:twocyltorque}. The profile
has to be continuous at $x=-d/2-\deltao$. This condition yields
\begin{eqnarray}
  \hc & = & - \sqrt{R^{2} - \deltao^{2}}
    + \frac{\deltao\lambda}{R} \exp{(-\deltao/\lambda)}
  \nonumber \\
  & = & - R + \frac{\lambda}{R} \,\deltao + \mathcal{O}(\deltao^{2})
  \nonumber \\
  & = & - R + \frac{\lambda\coth{\frac{d}{2\lambda}}}{R} \,\deltai+ \mathcal{O}(\deltai^{2})
  \; .
  \label{eq:equationforhc}
\end{eqnarray}
In the last step we exploited the fact that
$\deltao=\coth{(\frac{d}{2\lambda})}\,\deltai$ to first order
\cite{Weiklcyl}.

The profile of the membrane between the two cylinders in the
symmetric case is given by \cite{Weiklcyl}
\begin{equation}
  h_{\text{in}}(x) 
  = C + D \cosh{(x/\lambda)}
  \; ,
\end{equation}
where $C$ and $D$ are constants that can be determined from the conditions
of continuous profile and slope at $x=-d/2+\deltai$.
The latter condition yields
\begin{equation}
  D = \frac{\lambda \, \deltai}{\sqrt{R^{2} - \deltai^{2}}} \cdot
     \frac{1}{\sinh{(\frac{-\frac{d}{2} + \delta_{i}}{\lambda})}}
  = - \frac{\lambda\deltai}{R\sinh{(\frac{d}{2\lambda})}}+ \mathcal{O}(\deltai^{2})
  \; ,
  \label{eq:equationforD}
\end{equation}
the former
\begin{eqnarray}
  C & = & -\Big(\hc + \sqrt{R^{2} - \delta_{i}^{2}}\Big)
    - D \cosh{\Big(\frac{-\frac{d}{2} + \deltai}{\lambda}\Big)}
  \nonumber \\
  & \stackrel{(\ref{eq:equationforD})}{=} &
  -\hc - R +\frac{\lambda \coth{(\frac{d}{2\lambda})}}{R} \; \deltai
   + \mathcal{O}(\deltai^{2})
  \stackrel{(\ref{eq:equationforhc})}{=} \mathcal{O}(\deltai^{2})
  \; . \;\;\;\;\;\;\;\;\;
\end{eqnarray}
To first order the depth $\ho$ at the mid-line of the profile \cite{h0}
is thus given by
\begin{equation}
  \ho = -h_{\text{in}}(0) = -(C + D) =
  \frac{\lambda\deltai}{R\sinh{(\frac{d}{2\lambda})}}+ \mathcal{O}(\deltai^{2})
  \; .
\end{equation}
It can also be obtained from Eqn.~(\ref{eq:ytorquecylmemsym}). Using
$K_{\perp}= -h''_{\text{in}}(0)= -\frac{D}{\lambda^{2}}$ yields 
\begin{equation}
  \ho = -\frac{D + \frac{D^{2}}{2\lambda^{2}} \hc}
    {1 - \frac{D^{2}}{2\lambda^{2}} }
  =  \frac{\lambda\deltai}{R\sinh{(\frac{d}{2\lambda})}}+ \mathcal{O}(\deltai^{2})
  \; .
  \label{eq:hosmallgradientapproximation}
\end{equation}
The two results coincide at first order which confirms the validity 
of Eqn.~(\ref{eq:ytorquecylmemsym}) within this approximation.




\begin{thebibliography}{99}

\bibitem{Lodish}
H. Lodish, A. Berk, S. L. Zipursky, P. Matsudaira, D. Baltimore, and J. Darnell,
\emph{Molecular Cell Biology} (Freeman \& Company, New York, 2000).

\bibitem{HandbookofBioPhys}
\textit{Handbook of Biological Physics}, edited by R. Lipowsky and E. Sackmann
(Elsevier, Amsterdam, 1995) Vol. 1.

\bibitem{Marsh01}
M. Marsh (ed.), \emph{Endocytosis}, Frontiers in Molecular
Biology, (Oxford University Press, Oxford, 2001).

\bibitem{Robinson97}
M. S. Robinson, Trends Cell Biol. \textbf{7}, 99 (1997).

\bibitem{McMahonGallop05}
H. T. McMahon and J. L. Gallop, Nature \textbf{438}, 590 (2005).

\bibitem{Antonny06}
B. Antonny, Curr. Opinion Struc. Biol. \textbf{18}, 386 (2006).

\bibitem{LuHi92}
E. J. Luna, A. L. Hitt, Science \textbf{258}, 955 (1992).

\bibitem{Seifert97}
U. Seifert,
Adv. Phys. \textbf{46}, 13 (1997).


\bibitem{surfacestresstensor}
R. Capovilla and J. Guven,
J. Phys. A: Math. Gen. \textbf{35}, 6233 (2002).

\bibitem{Guven04}
J. Guven,
J. Phys. A: Math. Gen. \textbf{37}, L313 (2004).

\bibitem{Lomholt}
M. A. Lomholt and L. Miao,
J. Phys. A: Math. Gen. \textbf{39}, 10323 (2006).

\bibitem{Kozlov}
M. M. Kozlov,
J. Phys.: Condens. Matter \textbf{18}, S1177 (2006).


\bibitem{mem_inter}
M. M. M{\"u}ller, M. Deserno, and J. Guven,
Europhys. Lett. \textbf{69}, 482 (2005).

\bibitem{geomsurf}
M. M. M{\"u}ller, M. Deserno, and J. Guven,
Phys. Rev. E \textbf{72}, 061407 (2005).


\bibitem{Oettel}
A. Dominguez, M. Oettel,  and S. Dietrich,
arXiv:cond-mat/0611329 (2006).





\bibitem{Kralchevsky}
P. A. Kralchevsky, V. N. Paunov, N. D. Denkov, K. Nagayama,
J. Chem. Soc. Faraday Trans. \textbf{91}, 3415 (1995);
%
P. A. Kralchewsky and K. Nagayama, Adv. Coll. Interface
Sci. \textbf{85}, 145 (2000).

\bibitem{Koltover}
I. Koltover, J. O. R{\"a}dler, and C. R. Safinya,
Phys. Rev. Lett. \textbf{82}, 1991 (1999).

\bibitem{gbp}
M. Goulian, R. Bruinsma, and P. Pincus,
Europhys. Lett. \textbf{22}, 145 (1993);
%
Erratum: Europhys. Lett. \textbf{23}, 155 (1993);
%
note also the further correction in:
J.-B. Fournier and P. G. Dommersnes,
Europhys. Lett. \textbf{39}, 681 (1997).

\bibitem{inclusions}
V. I. Marchenko and C. Misbah,
Eur. Phys. J. E \textbf{8}, 477 (2002);
%
D. Bartolo and J.-B. Fournier,
Eur. Phys. J. E \textbf{11}, 141 (2003).

\bibitem{conicalinclusions}
T. R. Weikl, M. M. Kozlov, and W. Helfrich,
Phys. Rev. E \textbf{57}, 6988 (1998).

\bibitem{Weiklcyl}
T. R. Weikl,
Eur. Phys. J. E \textbf{12}, 265 (2003).

\bibitem{Fourspher}
P. G. Dommersnes, J.-B. Fournier, and P. Galatola,
Europhys. Lett. \textbf{42}, 233 (1998).

\bibitem{Kim}
K. S. Kim, J. Neu, and G. Oster,
Biophys. J. \textbf{75}, 2274 (1998).

\bibitem{BisBis}
P. Biscari, F. Bisi, and R. Rosso,
J. Math. Biol. \textbf{45}, 37 (2002);
%
P. Biscari and F. Bisi,
Eur. Phys. J. E \textbf{7}, 381 (2002).


\bibitem{deGennes}
P.-G. de Gennes, F. Brochard-Wyart, and D. Quere,
\textit{Capillarity and Wetting Phenomena}, (Springer, Berlin, 2003).

\bibitem{Canham}
P. B. Canham,
J. Theoret. Biol. \textbf{26}, 61 (1970).

\bibitem{Helfrich}
W. Helfrich,
Z. Naturforsch. \textbf{28c}, 693 (1973).


\bibitem{Eulerelastica}
M. Nizette and A. Goriely,
J. Math. Phys. \textbf{40}, 2830 (1999).


\bibitem{DNA1}
M. Doi and S. F. Edwards, 
\emph{The Theory of Polymer Dynamics}, 
(Cambridge University Press, 1986), Sec. 8.8.

\bibitem{DNA2}
O. Kratky and G. Porod, 
Rec. Trav. Chim. \text{68}, 1106 (1949).

\bibitem{DNA3}
I. M. Kulic, Ph.D. thesis, University of Mainz 2004.



\bibitem{DifferentialGeometry}
For a detailed introduction to the differential geometry of
two-dimensional surfaces see:
M. Do Carmo,
\textit{Differential Geometry of Curves and Surfaces}, (Prentice Hall,
1976);
%
E. Kreyszig,
\textit{Differential Geometry}, (Dover, New York 1991).

\bibitem{Frankel}
T. Frankel,
\textit{The Geometry of Physics}, (Cambridge University Press,
2003).

\bibitem{FournierMonge}
J.-B. Fournier, Soft Matter (accepted); see also 
arXiv:cond-mat/0702279 (2007).

\bibitem{Stamou_Fournier}
D. Stamou, C. Duschl, and D. Johannsmann,
Phys. Rev. E \textbf{62}, 5263 (2000);
%
J.-B. Fournier and P. Galatola,
Phys. Rev. E \textbf{65}, 031601 (2002).

\bibitem{planarelastica}
G. Arreaga, R. Capovilla, C. Chryssomalakos, and J. Guven,
Phys. Rev. E \textbf{65}, 031801 (2002).

\bibitem{LandauLifschitz1}
L. D. Landau and E. M. Lifshitz,
\textit{Mechanics}, 3rd ed.
(Butterworth-Heinemann, Oxford, 1976).

\bibitem{shapeeqn_meniscus}
The shape of the interface between two incompressible 
fluids under gravity (with density difference $\Delta\rho>0$ 
and gravitational acceleration $g$)
is determined by the Young-Laplace law which states that 
the hydrostatic pressure $P=-\Delta\rho g h$ equals 
the interfacial tension $\sigma$ times the curvature $K$:
$\dot{\psi} = h/\ell^{2}$, with $\ell=\sqrt{\sigma/(g\Delta\rho)}$. 
By differentiating this equation and equating it with 
$\dot{h}=\sin{\psi}$ we obtain Eqn.~(\ref{eq:sineGordon}) where $\lambda=\ell$ 
and $\eta=1$.

\bibitem{negativeangles}
The other solution with $\alphao<0$ and $\dot{\psi}\ge0$ can be simply
obtained by a reflection on the $(x,y)$ plane.

\bibitem{etalower1sym}
The expression for $\eta<1$ in Eqn.~(\ref{eq:psiinnersectionsym}) 
is valid for 
$0\le\tilde{s}\le\frac{2}{\sqrt{|\eta|}}\operatorname{K}(\frac{1}{m})$. 

\bibitem{Abramowitz}
\textit{Handbook of Mathematical Functions},
9th ed.,
edited by M. Abramowitz and I. A. Stegun
(Dover, New York, 1970).

\bibitem{LaLi_elast}
L. D. Landau and E. M. Lifshitz,
\textit{Theory of Elasticity}, 3rd ed.
(Butterworth-Heinemann, Oxford, 1986).

\bibitem{Seifert90}
U. Seifert and R. Lipowsky, Phys. Rev. A \textbf{42}, 4768 (1990).

\bibitem{boundarycondition}
M. Deserno, M. M. M\"uller, and J. Guven, 
submitted. 

\bibitem{maximum}
The intermediate maximum is not considered in the following as
it does not correspond to a stable profile.

\bibitem{membranesimulations}
B. J. Reynwar \etal, accepted.

\bibitem{h0}
Note that $\ho$ is defined differently in Ref.~\cite{Weiklcyl};
there it is given by $-(\hc + R)$.





\end{thebibliography}
\end{document}